# Light-induced anomalous Hall effect in graphene

J.W. McIver[1*], B. Schulte[1*], F.-U. Stein[1*], T. Matsuyama[1], G. Jotzu[1], G. Meier[1] and A. Cavalleri[1,2]

**Many striking non-equilibrium phenomena have been discovered or predicted in optically-driven quantum solids[1], ranging from light-induced superconductivity[2,3] to Floquet-engineered topological phases[4-8]. These effects are expected to lead to dramatic changes in electrical transport, but can only be comprehensively characterized or functionalized with a direct interface to electrical devices that operate at ultrafast speeds[1-8]. Here, we make use of laser-triggered photoconductive switches[9] to measure the ultrafast transport properties of monolayer graphene, driven by a mid-infrared femtosecond pulse of circularly polarized light. The goal of this experiment is to probe the transport signatures of a predicted light-induced topological band structure in graphene[4,5], similar to the one originally proposed by Haldane[10]. We report the observation of an anomalous Hall effect in the absence of an applied magnetic field. We also extract quantitative properties of the non-equilibrium state. The dependence of the effect on a gate potential used to tune the Fermi level reveals multiple features that reflect the effective band structure expected from Floquet theory. This includes a ~60 meV wide conductance plateau centered at the Dirac point, where a gap of approximately equal magnitude is expected to open. We also find that when the Fermi level lies within this plateau, the estimated anomalous Hall conductance saturates around ~1.8±0.4 $e^2/h$.**

Optical driving has been proposed as a means to engineer topological properties in topologically trivial systems[4-8]. One proposal for such a 'Floquet topological insulator' is based on breaking time-reversal symmetry in graphene through a coherent interaction with circularly polarized light[4]. In this theory, the light field drives electrons in circular trajectories through the band structure (Fig. 1a). Close to the Dirac point, these states are predicted to acquire a non-adiabatic Berry phase every optical cycle, which is equal and opposite for the upper and lower band. This time-averaged extra phase accumulation amounts to an energy shift that would lift the degeneracy of the Dirac point, opening a topological gap in the effective Floquet band structure (Fig. 1b).

* These authors contributed equally to this work
[1] Max Planck Institute for the Structure and Dynamics of Matter, Hamburg, Germany
[2] Department of Physics, Clarendon Laboratory, University of Oxford, Oxford, UK
  email: andrea.cavalleri@mpsd.mpg.de; james.mciver@mpsd.mpg.de

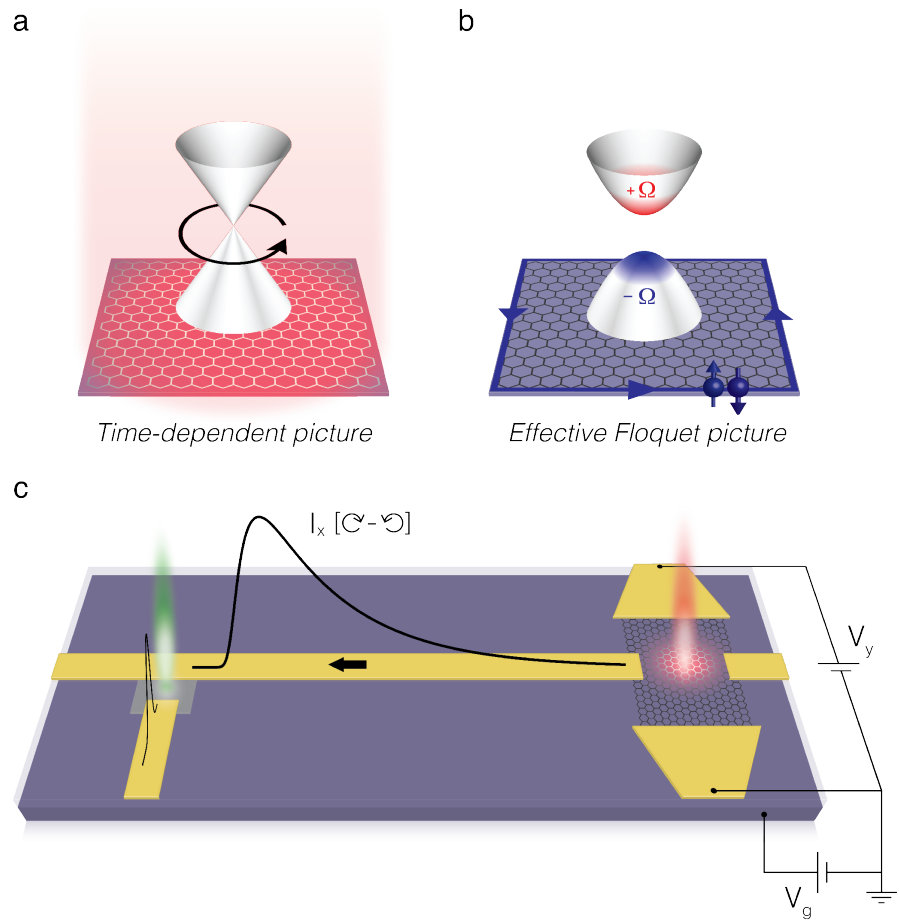

**Fig. 1 | Light-induced topological Floquet bands in graphene and device architecture used to detect ultrafast anomalous Hall currents. a,** A coherent interaction between graphene and circularly polarized light is predicted to open a topological band gap in the effective Floquet band dispersion[4] **(b)**. The gap is characterized by the presence of Berry curvature ($\Omega$), which is identical in both valleys. The experimental signature of the induced nontrivial topology is the emergence of anomalous Hall currents. **c,** Exfoliated graphene monolayer with four electrical contacts (right) and a photoconductive switch for current detection (left), connected by a microstrip transmission line. The graphene was optically driven using an ultrafast mid-infrared circularly polarized laser pulse (red beam). The generated helicity-dependent anomalous Hall currents $I_x$ [↻ - ↺] were probed after a variable time delay at the photoconductive switch, which was activated by a second laser pulse (green beam). Anomalous Hall currents were measured as a function of source-drain voltage bias $V_y$ and backgate voltage $V_g$, the latter of which controlled the graphene Fermi level ($E_F$).

The non-trivial topology of the Floquet bands forming this gap arises from their nonzero Berry curvature distribution[4,5], which integrated over the Brillouin zone defines a topological invariant, called the Chern number[11-14]. Topologically protected transport is predicted to develop if the Fermi level ($E_F$) lies inside the gap, exhibiting a quantized anomalous Hall conductance of $2e^2/h$ carried by edge states in the absence of an applied magnetic field[5,10-15]. This corresponds to the formation of a light-induced Chern insulator, equivalent to the phase originally proposed by Haldane[10] and distinct from topological phases induced by spin-orbit interaction[12-14,16,17]. While quantum simulation experiments have validated aspects of this proposal in synthetic physical settings[18,19], and Floquet-Bloch states have been detected in a topological insulator[20], anomalous Hall currents originating from such a light-induced topological band structure have not been observed in a real material.

What makes this proposal unique is that the anomalous Hall effect arises from Berry curvature that is coherently induced by light in a material where none is present in equilibrium. This is in contrast to previous observations of photo-induced anomalous Hall effects in semiconductor quantum wells[21], monolayer transition metal dichalcogenides[22] or Weyl semimetals[23], which originate from Berry curvature intrinsic to the equilibrium band structure due to broken inversion symmetry.

Inducing and detecting anomalous Hall currents in graphene presents multiple experimental challenges. The laser electric field strength required to open an observable topological gap is estimated to be $\sim 10^7\text{-}10^8$ V/m, even at mid-infrared wavelengths where the effect is enhanced[4,5,15]. Hence, to avoid material damage while still providing sufficient field strength, ultrafast laser pulses must be used. Consequently, the resulting Hall conductance changes are too short-lived to be probed with conventional transport techniques.

In this work, ultrafast anomalous Hall currents were detected on-chip by using a laser-triggered photoconductive switch[9]. A schematic of our device architecture is shown in Fig. 1c. An exfoliated monolayer graphene flake was transferred onto a doped silicon wafer with an oxide layer and contacted in a four-probe Hall geometry using standard lithography procedures[24,25]. The metallic leads formed microstrip transmission lines in conjunction with the oxide layer and silicon wafer. These routed ultrafast anomalous Hall currents generated in the graphene to a photoconductive switch for detection. The switch consisted of a resistive amorphous silicon patch that bridged the main transmission line and a probing line. When excited with a visible ultrafast laser pulse, the switch became highly conductive and detected currents flowing in the main transmission line with a time resolution set by the silicon carrier lifetime ($\sim 1$ ps). By adjusting the time delay between the graphene laser drive pulse (pump) and the switch trigger pulse (probe), the temporal profile of ultrafast anomalous Hall currents could be characterized. The amplitude of the detected currents were determined by calibrating the photovoltaic response of the switch using a DC bias field, which is possible in our microstrip geometry because only the fundamental quasi-TEM mode propagates up to ∼ THz frequencies.

Graphene was driven using a ∼ 500 fs laser pulse at a frequency of ∼ 46 THz (photon energy $\hbar\omega$ ∼ 191 meV, wavelength ∼ 6.5 $\mu$m). Unless otherwise noted, a peak laser pulse fluence of ∼ 0.23 mJ/cm$^2$ was used, corresponding to a peak intensity of ∼ 4.3 x 10$^{12}$ W/m$^2$ and peak electric field strength of ∼ 4.0 x 10$^7$ V/m (in free space, for circular polarization). The pulses were focused to a spot size of ∼ 80 $\mu$m (FWHM), ensuring homogeneous illumination of the graphene flake and the contacts. A second ultrafast laser pulse centered at 520 nm was used to operate the photoconductive switch. The device was mounted in a microscopy cryostat designed for high-frequency transport measurements and cooled to a base temperature of 80 K. A global backgate formed by the silicon wafer and the oxide layer controlled the Fermi level ($E_F$) in the graphene flake. The graphene field-effect mobility was measured to be $\mu$ ∼ 10,000 cm$^2$/Vs in the vicinity of the Dirac point. The results presented here are from a single device, and consistent results have been obtained using five different devices.

Anomalous Hall currents induced by circularly polarized light are expected to exhibit the following traits: (1) They should be generated in the transverse direction ($I_x$) with respect to an applied DC voltage bias ($V_y$). (2) They should reverse polarity upon reversing the light helicity. (3) They should reverse polarity upon reversing $V_y$, with a linear functional dependence. To probe (1) and (2), we directly detected the difference between currents $I_x$ generated with right- versus left-circular polarization (henceforth referred to as $I_x$ [↻ - ↺]), utilizing an optical polarization chopping technique.

Figure 2a displays the measured $I_x$ [↻ - ↺] signal as a function of pump-probe time delay for a positive and negative $V_y$, with the Fermi level gated to the equilibrium graphene Dirac point ($E_F$ = 0). The time-resolved signal exhibits a fast rise time followed by an exponential decay, and reverses polarity upon reversing $V_y$.

Note that the data in Fig. 2a do not directly reflect the timescale on which the anomalous Hall effect was generated in the graphene. The signal was transformed as it propagated through the on-chip circuitry. We modelled these effects through a combination of simulations and direct calibrations of the system parameters including the contact resistance, graphene capacitance, microstrip impedance and dispersion in the transmission line (see supplementary S3). The reconstructed current profiles are shown in Fig. 2b.

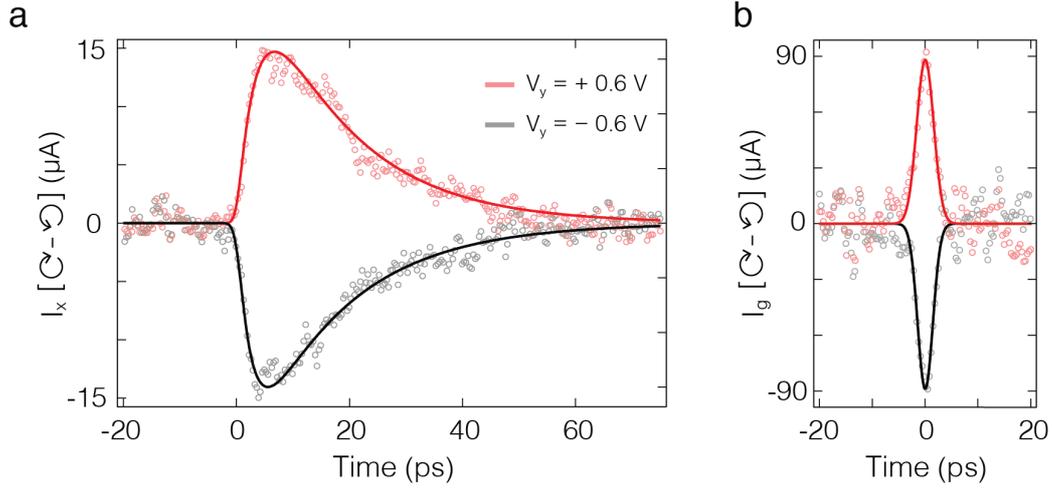

**Fig. 2 | Ultrafast anomalous Hall currents in graphene driven by circularly polarized light.**
**a**, Time-resolved helicity-dependent anomalous Hall currents $I_x$ [↻ - ↺] measured at the photoconductive switch for a positive (red) and negative (black) transverse source-drain voltage $V_y$. The graphene Fermi level was gated to the Dirac point ($E_F = 0$). Solid lines are based on the signal propagation model in supplementary S3. A small background observed at $V_y = 0$ was subtracted from the data sets (see supplementary S4). **b**, Reconstructed anomalous Hall current signals $I_g$ accounting for the response function of the on-chip circuitry. The current amplitude has a ±22% systematic error from the calibration of the photoconductive switch. Solids lines are Gaussian fits.

Having established the presence of ultrafast anomalous Hall currents in graphene, we investigated the functional dependence of $I_x$ [↻ - ↺] on $V_y$ at $E_F = 0$. We did so by fixing the pump-probe time delay at the maximum of the $I_x$ [↻ - ↺] signal in Fig. 2a, which we refer to as $\hat{I}_x$ [↻ - ↺], and measured the signal amplitude as a function of $V_y$ (Fig. 3a). The data exhibit the expected linear dependence. We also investigated helicity-dependent currents generated in response to a longitudinal voltage bias $V_x$ (Fig. 3b), which should be forbidden by symmetry. The data in Fig. 3b confirm that no helicity-dependent longitudinal currents were generated.

We define the peak anomalous Hall conductance as $G_{xy} = \hat{I}_g$ [↻ - ↺] $/ 2V_y$, where $\hat{I}_g$ is the peak of the reconstructed signal in Fig. 2b. Figure 4a displays $G_{xy}$ as a function of the laser drive pulse fluence, measured for $E_F = 0$. At the highest achievable fluence, we estimate $G_{xy} \cong (1.8 \pm 0.4)\ e^2/h$, consistent with recent numerical simulations of optically-driven graphene for our laser pulse parameters[26].

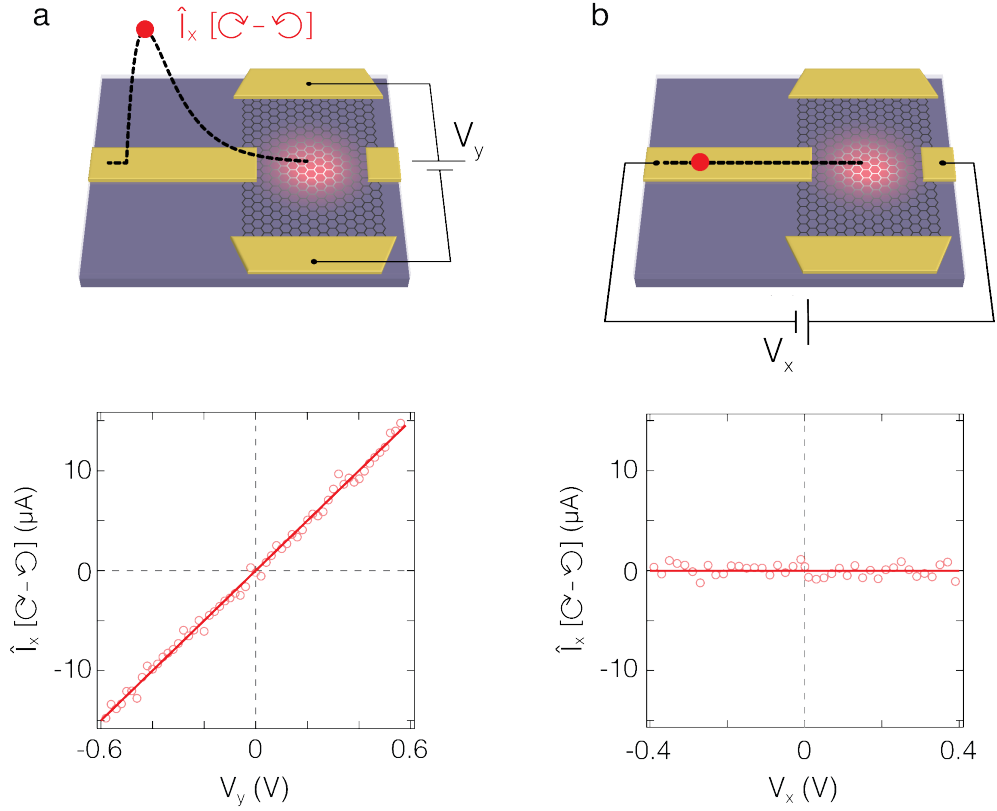

**Fig. 3 | Helicity-dependent current behavior under different source-drain voltage geometries.** The peak of the helicity-dependent current $\hat{I}_x$ [↻ - ↺] measured as a function of **a**, transverse source-drain voltage $V_y$ and **b**, longitudinal source-drain voltage $V_x$. The graphene Fermi level was gated to the Dirac point ($E_F = 0$). Solids lines are linear fits.

Multiple theories have predicted $G_{xy} \approx 2e^2/h$ to result from the emergence of topological Floquet bands in graphene, provided that $E_F$ lies inside the gap that opens at the Dirac point[5,15,27-29]. To characterize the predicted topological bands, we varied the equilibrium $E_F$ using the backgate[30] and compared the variation in $G_{xy}$ (Fig. 4b) with calculations of the expected effective band structures based on Floquet theory for our laser pulse parameters (Fig. 4c) (see supplementary S7).

For low drive fluence (red circles), $G_{xy}$ was observed to be independent of the backgate potential for $|E_F| \lesssim \hbar\omega/2$, where direct interband transitions are allowed. This suggests that as the predicted gaps $\Delta_1$ open at $\pm\hbar\omega/2$, the resulting electron distribution in the Floquet bands, in conjunction with the Berry curvature near these gaps[4,15,27-29,31], acts as a source of anomalous Hall currents. As the fluence was increased (blue and black circles), features emerged near $\pm\hbar\omega/2$ that are closely aligned with the gaps $\Delta_1$. For $|E_F| \gtrsim \hbar\omega/2$, $G_{xy}$ decreased to zero and changed sign when approaching the dielectric breakdown threshold of the device.

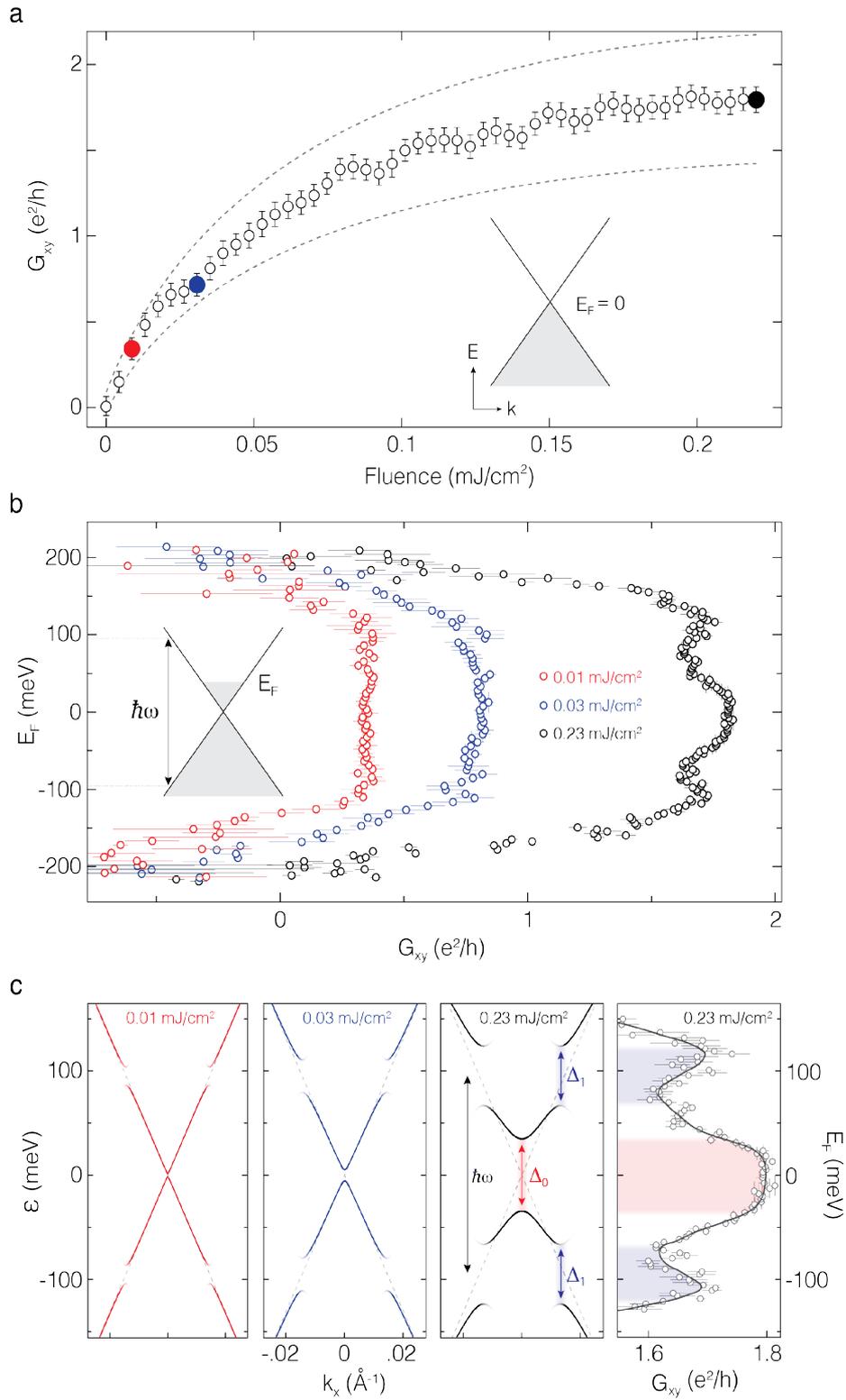

**Fig. 4 | Evidence for topological Floquet bands. a,** Anomalous Hall conductance $G_{xy}$ as a function of the peak laser drive pulse fluence. The equilibrium Fermi level was gated to the Dirac point ($E_F = 0$). Error bars are the standard statistical error, and the dashed lines denote the systematic error from the calibration of the photoconductive switch. Colored data points correspond to the fluences measured in b. **b,** $G_{xy}$ as a function of $E_F$ measured at three fluences. Horizontal error bars are the standard error and vertical error bars denote the uncertainty related to determining the precise value of the Dirac point. The systematic error on $G_{xy}$ is the same as in (a). **c,** Left three panels: Effective band structures for the fluences reported in (b) simulated using Floquet theory. At the highest fluence, we calculated $\Delta_0 \cong 69$ meV and $\Delta_1 \cong 56$ meV. Right panel: Blow up of the high-fluence data in (b) for comparison. Solid line is the smoothed data. Shaded regions highlight the features corresponding to light-induced band gaps in the Floquet band structure.

At the highest fluence (black circles), close to the Dirac point and away from the resonance, a conductance plateau was observed for $|E_F| \lesssim 30$ meV, where $G_{xy} \cong (1.8 \pm 0.4)$ $e^2/h$ (Fig. 4c right). Remarkably, the width of the plateau is very close to the calculated width of the light-induced topological gap at the Dirac point $\Delta_0$ (69 meV, red shading). For $|E_F| \gtrsim \Delta_0/2$, $G_{xy}$ decreased, as predicted due to the reduced net Berry curvature of the occupied dressed electronic states[11-14]. The sharpness of this decrease may depend on multiple factors, including the Berry curvature and non-equilibrium electron distributions in the Floquet bands, which is influenced by the frequency, amplitude and pulse-shape of the laser, as well as disorder, interactions and dissipation[4,15,27-29,31].

In summary, we observed an ultrafast light-induced anomalous Hall effect in graphene that displayed multiple signatures of emergent topological Floquet bands. Experimental advances utilizing encapsulated graphene heterostructures will open up the regime $|E_F| \gg \hbar\omega/2$, where $G_{xy}$ reverses sign, in future experiments. Our studies may also be extended to other compelling problems in quantum materials, including the non-equilibrium control of van der Waals heterostructures such as twisted bilayer graphene[32], transition metal dichalcogenides[33], materials undergoing photo-induced phase transitions[34], and the physics of non-equilibrium superconductors[2-3].

Supplementary information for

# Light-induced anomalous Hall effect in graphene

J.W. McIver[*], B. Schulte[*], F.-U. Stein[*], T. Matsuyama, G. Jotzu, G. Meier and A. Cavalleri

**Table of contents**

# S1. Experimental setup and methods

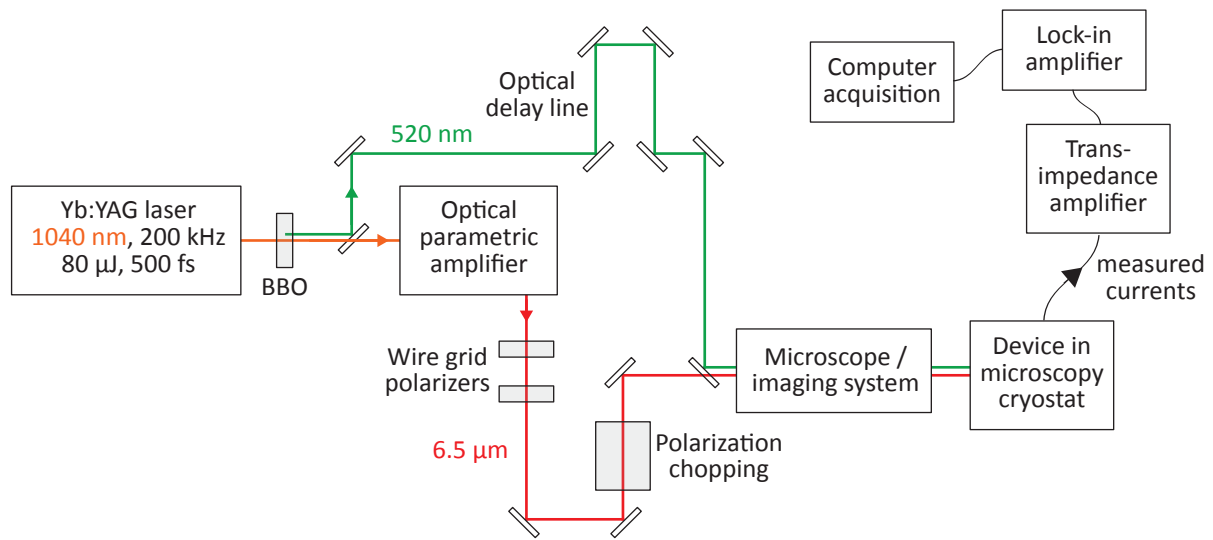

**Fig. S1 | Block diagram of the experimental setup.**

## S1.1: Experimental setup

**Setup overview.** A block diagram of the experimental setup used to induce and probe ultrafast anomalous Hall currents in graphene is depicted in Fig. S1. A train of ultrafast laser pulses with a central wavelength of 1040 nm, pulse duration of ∼ 500 fs and pulse energy of ∼ 80 μJ were derived from a commercial Yb-based laser system. It operated at 211 kHz with an average power of 16 W. Approximately 1% of the total power was converted to the second harmonic wavelength of 520 nm by passing the beam through a Beta Barium Borate (BBO) crystal. The remaining 1040 nm light was separated from the 520 nm light using a dichroic mirror and fed into a home-built optical parametric amplifier[35] with a difference frequency generation stage. This produced mid-infrared laser pulses at 6.5 μm with a pulse duration of ∼ 500 fs.

The mid-infrared beam was passed through two wire-grid polarizers, which were used to adjust the intensity and linear polarization of the light. A quarter-waveplate mounted in a hollow-bore motor was used to perform helicity-dependent measurements using a polarization chopping technique (discussed below). From there the beam was coupled into a microscope and focused onto the graphene device at normal incidence with a ∼ 80 μm focused spot size (FWHM) so that the graphene and contacts were homogeneously illuminated. The device was mounted in a microscopy cryostat designed for high frequency transport measurements and cooled to a base temperature of 80 K for all measurements.

The 520 nm laser beam generated at the output of the laser was routed through an optical delay line, coupled into the microscope and used to operate the photoconductive switch using a peak fluence of ~ 15 mJ/cm$^2$. A microscope camera provided a live video feed of the laser beam and device alignment, which was maintained using active stabilization techniques. By adjusting the path length of the optical delay line, the timing between the 6.5 μm graphene optical drive pulse (pump) and the 520 nm switch trigger pulse (probe) could be varied and ultrafast anomalous Hall currents detected with a ~ 1 ps time resolution. Detected currents were amplified using a home-built transimpedance amplifier and processed using standard lock-in techniques.

**Optical polarization chopping.** We isolated helicity-dependent currents using optical polarization chopping techniques[36]. A hollow-bore motor was used to continuously rotate a quarter-waveplate at a frequency of ~ 200 Hz. The mid-infrared polarization then changed helicity four times per 360-degree revolution. By triggering on the 2$^{nd}$ harmonic of the rotation frequency, helicity-dependent signals could be isolated using a lock-in amplifier. We calibrated the phase of the lock-in amplifier by performing optical polarization measurements at the sample position.

**Intensity dependence data.** As discussed in the main text, we define the peak anomalous Hall conductance as $G_{xy} = \hat{I}_g[\circlearrowright - \circlearrowleft] / 2V_y$, where $\hat{I}_g \cong 6.0\,\hat{I}_x$ (see section S3.4). The data in Fig. 4a (main text) is the average of $G_{xy} = 6\,\hat{I}_x[\circlearrowright - \circlearrowleft] / 2V_y$ measured with $V_y$ = +0.6 V and $V_y$ = -0.6 V at $E_F$ = 0 meV.

**Backgate dependence data.** The data in Fig. 4b (main text) is the average of $G_{xy} = 6\,\hat{I}_x[\circlearrowright - \circlearrowleft] / 2V_y$ measured with a positive and negative $V_y$ (see section S3.4). Measurements were performed by applying +/- 12.5 V over a 100 kΩ resistor in series with the source so that the source-drain current running through the graphene was held approximately constant at ~ 125 μA for all gate voltages. This was done to prevent large source-drain currents from breaking the contacts as the graphene carrier density was increased by the gate. While measuring $\hat{I}_x[\circlearrowright - \circlearrowleft]$ at different backgate voltages, the current outputted by the source was simultaneously recorded and the DC source-drain voltage drop $V_y$ could be determined and used to calculate $\hat{I}_x[\circlearrowright - \circlearrowleft] / 2V_y$.

The equilibrium Fermi level position was calculated using the relation $E_F = \text{sign}(n)\,\hbar\,v_f\sqrt{\pi\,|n|}$, after Ref. 30, where $v_f$ is the Fermi velocity (we use $v_f = 1 \cdot 10^6$ ms$^{-1}$), $n = C\,(V - V_D)\,/\,e$ is the graphene carrier density, $C = \varepsilon_0\varepsilon_r\,/\,d$ is the capacitance per area of the graphene/SiO$_2$ stack, $e$ is the electron charge, $V_D$ is the Dirac point, $V$ is the applied gate voltage, the thickness $d$ = 285 nm and $\varepsilon_r$ = 3.9. The transfer characteristic is shown in Fig. S3e.

## S1.2: Beam characterization

Mid-infrared laser pulses derived from the OPA were collimated, passed though wire-grid polarizers, and a quarter-waveplate. A 10x reflective microscope objective was used for imaging and to focus the mid-infrared light onto the graphene flake. The beam entering the objective was imaged using a Pyrocam Beam Profiling Camera from Ophir. A picture of the round Gaussian profile is shown as an inset in Fig. S2a.

The beam size at the focus position, at the plane of the graphene flake, was determined by standard knife edge measurements. A razor blade was mounted on a translation stage and moved into focus behind the microscope objective. The power was recorded behind the razor blade for different positions. The result for the *x*-direction (parallel to the table surface) and *y*-direction (vertical to the table) is shown in Fig. S2. We find a beam profile with FWHM$_x$ = 83 µm and FWHM$_y$ = 91 µm, corresponding to Gaussian widths $\sigma_x$ = 35.5 µm and $\sigma_y$ = 39 µm. The temporal FWHM of the mid-infrared laser pulses is 500 fs, which yields a width of $\sigma_t$ = 212 fs. With these parameters we can determine the peak fluence reported in the main text and the peak electric used for Floquet calculations (see supplementary S7).

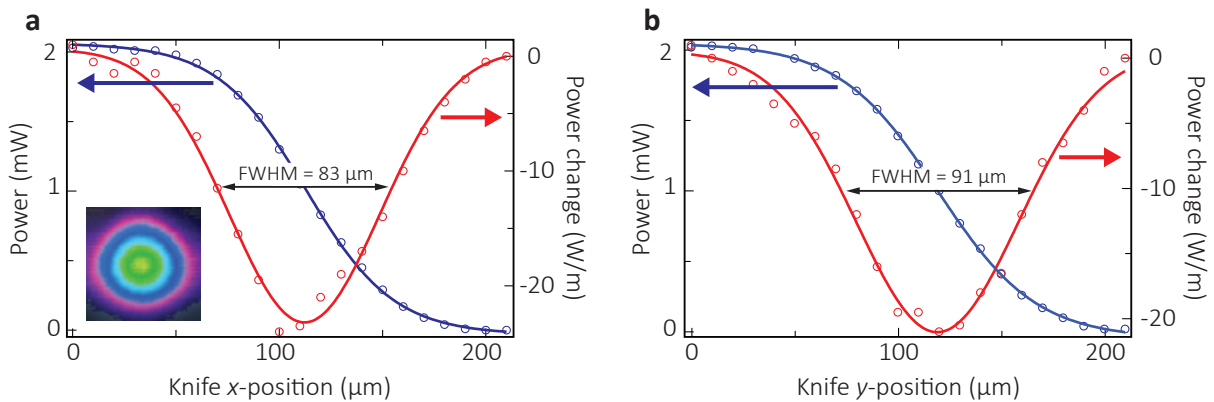

**Fig. S2 | Beam profile.** A razor blade was moved into the beam at the position of the focus behind the reflective microscope objective and the total transmitted power passing the knife edge was measured. **a,** Transmitted power (blue, left axis) and differentiated power (red, right axis) measured for different knife edge position along the x-axis (parallel to the table). The inset show the beam profile before it is focused by the reflective microscope objective. **b,** Transmitted power (blue, left axis) and differentiated power (red, right axis) measured for different knife edge position along the y-axis (vertical to the table).

The peak fluence $F_{max}$ for a given average power $P$ behind the diamond windows of the cryostat is given by:

$$F_{max}(P) = \frac{P}{f_{rep}} \cdot \underbrace{\frac{1}{2\pi\,\sigma_x\sigma_y}}_{\text{space integral}} \;,$$

where $f_{rep}$ = 211 kHz is the repetition of the laser.

The peak electric field in the center of the beam in free space for linear polarization ($E_{max,lin}$) and circular polarization ($E_{max,circ}$) can be calculated by

$$E_{max,lin}(P) = \sqrt{2}\,E_{max,circ}(P) = \sqrt{\frac{2}{\varepsilon_0 c} \cdot \frac{P}{f_{rep}} \cdot \underbrace{\frac{1}{\sqrt{2\pi}\,\sigma_t}}_{\text{time integral}} \cdot \underbrace{\frac{1}{2\pi\,\sigma_x\sigma_y}}_{\text{space integral}}} \;,$$

where $c$ is the speed of light and $\varepsilon_0$ is the vacuum permittivity.

Note that the peak electric field is reduced at the graphene position compared to the free space value due to interference between the incident light and reflected light from the underlying silicon substrate. Assuming no shielding from the graphene flake itself, we modelled this field reduction by considering the substrate as a stack consisting of a 285 nm SiO$_2$ layer and a semi-infinite slab of highly doped p-type silicon, with refractive indices at 6.5 µm wavelength of 1.2 and 3.08 + 0.152 $i$ (Ref. 37), respectively. We calculated that the peak field at the location of the graphene flake was reduced by ∼ 36 % compared to the field in free space.

# S2. Device fabrication and characterization

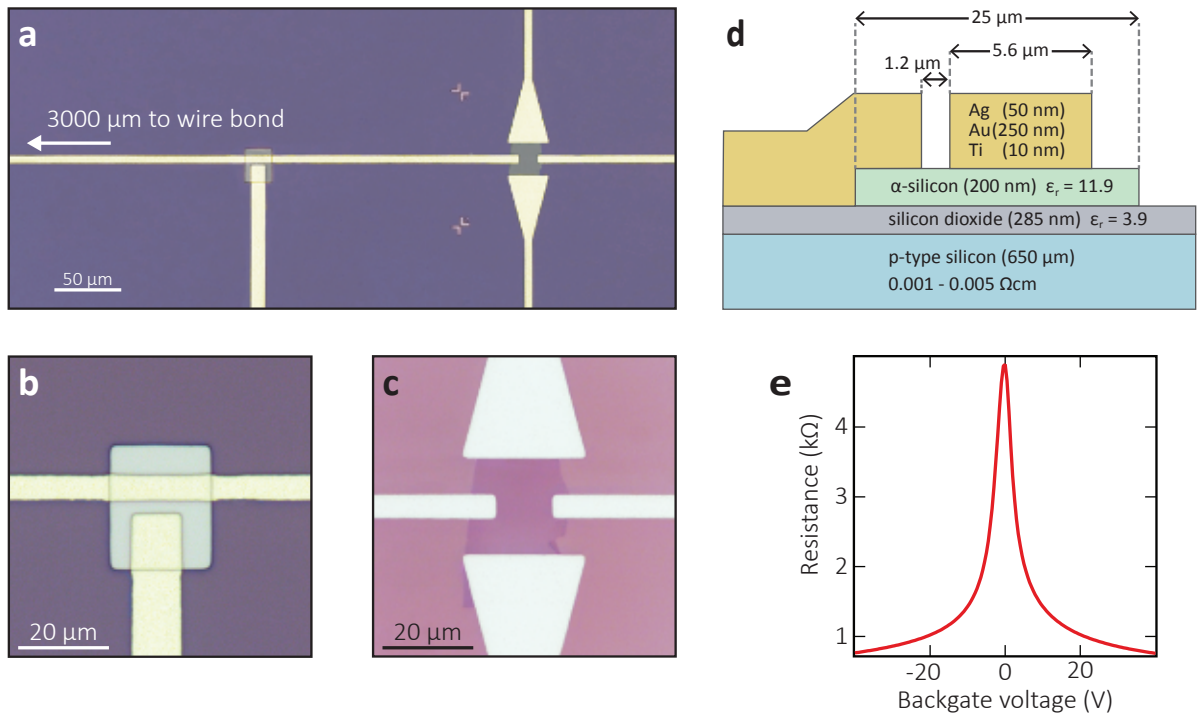

**Fig. S3 | Sample geometry and transport characteristics. a**, Optical microscopy image of the device studied in the main text. **b**, Close-up of the silicon photoconductive switch. **c**, Close-up of the contacted graphene flake. **d**, Schematic cross-section at the photoconductive switch position (not to scale). **e**, Graphene transfer characteristic: two-point resistance measurement between the large graphene contacts as a function of backgate voltage at a temperature of 80 K. The graphene charge neutrality point remained at a backgate voltage between -1 V and +1 V for all measurements reported in the paper.

Single layer graphene was obtained by mechanical exfoliation. Subsequently, the ultrafast optoelectronic circuitry was fabricated using standard laser lithography, thermal evaporation, and lift-off processing techniques.

Figure S3a-c shows microscope images of the graphene device studied in the main text. Graphene was mechanically exfoliated from Kish graphite by the tape method[24] and transferred onto a highly doped p-type silicon wafer with a 285 nm $SiO_2$ layer. The investigated single layer graphene flake was identified by optical contrast imaging and Raman spectroscopy. After choosing the flake, the optoelectronic circuitry was fabricated in two steps. First, an amorphous silicon ($\alpha$-silicon) patch was thermally evaporated in close proximity to the graphene flake. This formed the base of the photoconductive switch (Fig. S3b). Second, layered Ti/Au/Ag metallic structures were deposited to create signal lines.

These formed microstrip transmission lines in conjunction with the SiO$_2$ layer and the doped silicon wafer that were capable of transporting ultrafast currents. Four such signal lines were contacted to the graphene flake (Fig. S3c). One of these lines also passed over the α-silicon patch. The α-silicon patch was contacted with an additional sampling line so that a narrow semiconducting channel was formed between the main signal line and the sampling line (Fig. S3b). Together this structure formed the photoconductive switch. A schematic cross-section at the photoconductive switch position is shown in Fig. S3d. The circuitry is discussed in detail in the next section.

For both steps a bilayer photoresist system was used. The bottom layer was a heat-resistant undercut layer of *MicroChem LOR-7B*. The second layer consisted of positive photoresist *micro resist map-1205*. The structures were written with a direct-write laser lithography system. Afterwards the structures were developed and material was evaporated via thermal evaporation. The prepared chip was glued onto a printed circuit board chip carrier and contacted with wire bonds. To perform DC transport and optoelectronic measurements, the chip carrier was loaded into an optical microscopy cryostat equipped with low noise cables that connected the device to measurement electronics outside the cryostat.

The silicon substrate with an oxide layer was used as a global backgate to control the graphene Fermi level. A two-point resistance measurement at 80 K as a function of backgate voltage is shown in Fig. S3e. After spending time in vacuum and laser annealing with 6.5 μm light, the charge neutrality point settled very close to 0 V backgate voltage and remained between +1 V and -1 V for all measurements reported in the paper. The field-effect carrier mobility was determined to be μ ∼ 10,000 cm$^2$/Vs at a temperature of ∼ 80K.

# S3. Ultrafast circuitry

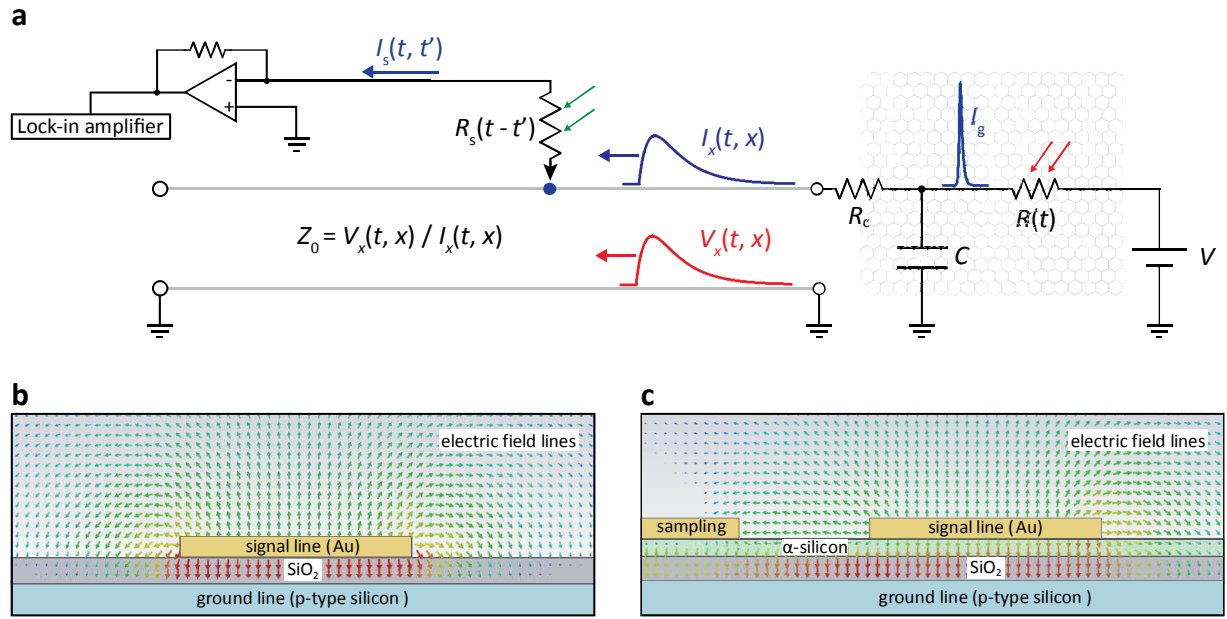

**Fig. S4 | Equivalent circuit model of ultrafast current generation, propagation and detection. a**, A current pulse $I_g$ is generated in the graphene, which is sourced from a voltage biased light-controlled resistor $R(t)$. The graphene capacitance $C$ has the effect of slowing down the discharge of the generated current into the microstrip transmission line with a time constant $\tau_{RC} \approx (Z_0 + R_c) C$, where $Z_0$ is the transmission line impedance and $R_c$ is the contact resistance. $I_x$ is the discharged current. At the position $x_s$ the photoconductive switch (excited at time $t'$) is modeled by a light-controlled resistor $R_s(t - t')$ that probes the local voltage profile $V_x$ of the current pulse in the transmission line. **b**, Simulation of the electric field line distribution in the transmission line at a fixed point in space and time (cross-sectional view). **c**, Electric field line simulation at the switch position. The field lines bend in the direction of the sampling line and create a current when the switch is illuminated by laser light.

## S3.1: Equivalent circuit model

***Current generation.*** An equivalent circuit model of the optoelectronic device architecture used to generate and detect ultrafast anomalous Hall currents in graphene is shown in Fig. S4a. The source of the current generation is modeled as a light-controlled adjustable resistor $R(t)$, representing the graphene anomalous Hall resistance, in series with a DC voltage bias $V$. Before the light pulse arrives, the anomalous Hall conductance is zero corresponding to $R = \infty$ (open circuit). When the light is on, the anomalous Hall conductance becomes finite and a current $I_g$ flows through the resistor due to the applied voltage bias. The source resistance $R(t)$ is large compared to the load $Z_0 + R_c$, which is given by the contact resistance $R_c \approx 260\ \Omega$ plus the transmission line impedance $Z_0 \sim 8.6\pm1.0\ \Omega$. The contact resistance was determined from 4-probe measurements.

The capacitive coupling between the graphene flake and the underlying doped Si substrate is non-negligible and can be modeled as a shunt capacitor $C \sim 37$ fF in parallel to the load $R_c + Z_0$. In conjunction with the load, the capacitor acts as a low-pass filter. During the light-on state, the current $I_g$ charges the capacitor. As the laser pulse subsides, $R \to \infty$ and $I_g \to 0$ and the capacitor discharges with a characteristic time constant $\tau_{RC} \sim (Z_0 + R_c) \cdot C$, which corresponds to the exponential decay of $I_x$. In reality, the finite equilibrium source resistance $R_{off} \sim 2$ k$\Omega$ of the graphene flake in the light-off-state has to be taken into account for the discharge process. When the source resistance in the light-off-state is high enough ($R_{off} \gg (Z_0 + R_c)$), then the majority of the photocurrent is discharged from the graphene. In other words, the graphene acts as an efficient current source. On p. 15ff, we show how $I_g$ can be recovered from $I_x$ using the transfer function for a low-pass filter.

***Signal propagation.*** The pulsed signal that is injected into the transmission line is characterized by a transient current component $I_x(t, x)$ and a voltage component $V_x(t, x)$ that vary in time and space. They are self-propagated by local AC electric and magnetic fields that are near-field confined to the transmission line. The fields form a quasi-TEM mode that moves with a group velocity $v = c / \sqrt{\varepsilon_{eff}} \sim c/2$ in the present experiment, where $\varepsilon_{eff} \sim 4$ is the effective dielectric constant of the transmission line geometry and $c$ is the speed of light. Since there are no travelling waves in any other directions, $V_x(t, x)$ and $I_x(t, x)$ are inevitably coupled at any position along the line by the characteristic impedance of the transmission line $Z_0 = V_x(t, x) / I_x(t, x) \sim 8.6\pm1.0\ \Omega$ (discussed in more detail in section S3.3).

The transmission line was optimized for propagating THz frequencies and designed so that $V_x(t, x)$ and $I_x(t, x)$ are in phase at all points in space and time. This was done using a combination of full wave analysis finite element simulations and high frequency lumped circuit simulations. Figures S4b-c plot simulations of the confined electric field lines of $V_x(t, x)$ at a fixed position in space and time for our transmission line design (cross-sectional view). The electric field is mainly confined underneath the Au microstrip. Thus, the photoconductive switch is only a small perturbation and possible reflections are negligible.

Note that as signals propagate in the transmission line they experience some dispersion. We determine the signal dispersion quantitatively for our transmission line geometry in section S3.2.

*Signal detection.* Our measurement directly probes the local electric fields of the microstrip associated with $V_x(t, x)$ at the position $x = x_s$ using a photoconductive switch. $x_s$ is approximately 185 µm relative to the graphene flake in the circuit design discussed in the main text. The switch can be modeled as a light-controlled resistor $R_s(t - t′)$, operated at time $t′$, that bridges the main signal line to a sampling line. As the signal passes by the switch, $V_x(t, x_s)$ biases the switch. Figure S4c shows the electric field line distribution in the transmission line at the position of the switch, where the field biasing can be seen as a vector field pointing from the signal line towards the sampling line. Without laser illumination, $R_s$ is highly resistive and has little influence on the signal as it passes by. Upon laser illumination, $R_s$ becomes highly conductive and a current $I_s(t, t′)$ flows into the contact of the sampling line due to the biasing voltage $V_x(t, x_s)$. By varying $t′$ the temporal profile of $V_x(t, x_s) = Z_0 \cdot I_x(t, x_s)$ can be sampled with $\sim$ 1 ps time resolution.

*Signal modelling.* In this section, we derive an analytical expression for $V_x(t, x)$ which is generated at the contact ($x = 0$) in the transmission line. In our model we take into account the capacitance $C$ of the graphene flake with the underlying substrate and the load, which is the characteristic impedance $Z_0$ of the transmission line plus the contact resistance $R_c$. A simple voltage source model with a DC-voltage bias $V$ and a time-dependent source resistance $R(t)$ (Fig. S4a) is used for modeling the current pulse launched from the graphene. With the safe assumption that the source resistance $R$ is much larger than the load impedance we obtain for the time-dependent voltage signal the convolution expression

$$V_x(t, x=0) \cong Z_0 \cdot I_g(t) * h_{\text{RC}}(t) \qquad (1)$$

with the impulse response function of the RC-system

$$h_{\text{RC}}(t) = \frac{1}{\tau_{\text{RC}}} e^{-\frac{t}{\tau_{\text{RC}}}} \cdot \theta(t) \qquad (2)$$

where $\tau_{\text{RC}} \approx (Z_0 + R_c) \cdot C$ is the characteristic RC time constant and $\theta(t)$ is the Heaviside step function. The impulse response function $h_{\text{RC}}(t)$ describes a first order low-pass characteristic with a cut-off frequency of $f_{RC}$ = 1 / (2π · $\tau_{RC}$) and results from the charging and discharging effect of the capacitance of the graphene flake. For later discussion we also introduce the transfer function in the frequency domain, which is given by:

$$V_{RC}(\omega) = \frac{1}{1 + i\tau_{RC}\omega} \quad (3)$$

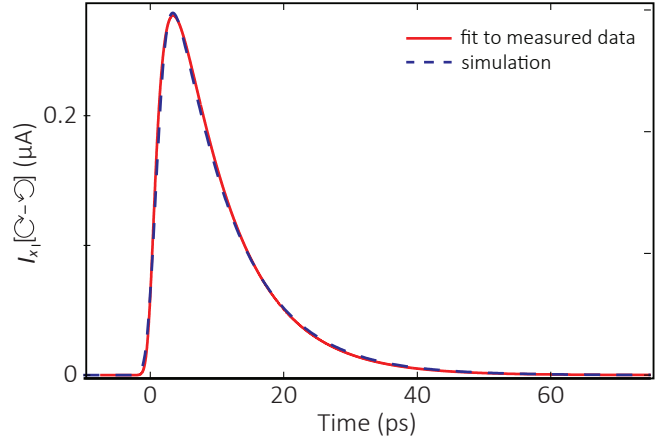

**Fig. S5 | Analytical modelling of the effective circuit.** Comparison of the analytical model (red, Eq. 5) fitted to dispersion corrected data (see section S3.2) with the lumped element circuit simulation (dashed blue). The capacitance is set to be $C$ = 37 fF, in accordance with the capacitance associated with the graphene flake on the $Si^{++}/SiO_2$ substrate. The decay time is $\tau_{RC} \approx (Z_0 + R_c) \cdot C$ = (8.6 Ω + 260 Ω)·37 fF = 10 ps. Note that dispersion corrected data is used in this simulation to reproduce the effect of the flake capacitance. The dispersion is characterized in the next section.

We model the temporal profile of the transient graphene resistance as a convolution of the Gaussian optical pulse profile (of width $\sigma$) used to photoexcite the graphene, and an intrinsic photocurrent exponential decay with a characteristic lifetime $\tau_g$. The resulting current can then be expressed as:

$$I_g(t) = \left[\frac{1}{\sqrt{2\pi}\sigma} e^{-\frac{t^2}{2\sigma^2}}\right] * \left[\frac{Q_g}{\tau_g} e^{-\frac{t}{\tau_g}} \cdot \theta(t)\right] \quad (4)$$

where $Q_g$ is the total charge in the generated pulse. Using Eq. 4 as an ansatz for the transfer function defined by Eq. 1 and the impulse response function in Eq. 2, we obtained an expression for the time-dependent voltage:

$$V_x(t,x=0) \cong \frac{Z_0 Q_g}{2(\tau_{RC} - \tau_g)} \left\{ e^{-\frac{t}{\tau_{RC}} + \frac{\sigma^2}{2\tau_{RC}^2}} \cdot \text{erfc}\left(\frac{1}{\sqrt{2}}\left(\frac{\sigma}{\tau_{RC}} - \frac{t}{\sigma}\right)\right) \right.$$
$$\left. - e^{-\frac{t}{\tau_g} + \frac{\sigma^2}{2\tau_g^2}} \cdot \text{erfc}\left(\frac{1}{\sqrt{2}}\left(\frac{\sigma}{\tau_g} - \frac{t}{\sigma}\right)\right) \right\} \quad (5)$$

Here erfc(*x*) is the complementary error function erfc(*x*) = 1-erf(*x*). This equation was used to fit the data in Fig. 2a (main text). For the given experiment we can safely assume that $\tau_{RC} \gg \tau_g$.

We compare the analytical model of Eq. 5 with lumped element circuit simulations. The circuit diagram shown in Fig. S4a is simulated and the current profile as a function of time is recorded. The results are shown in Fig. S5. Both describe a short pulse that is generated in the graphene flake and later discharged from the capacitance of the circuitry into the load represented by the contact resistance and the microstrip. We find that the lumped element circuit simulations using this model also accurately reproduce the more complex signals measured in other device geometries (see supplementary chapter S5).

## S3.2: Photoconductive switch response, signal dispersion and signal reconstruction

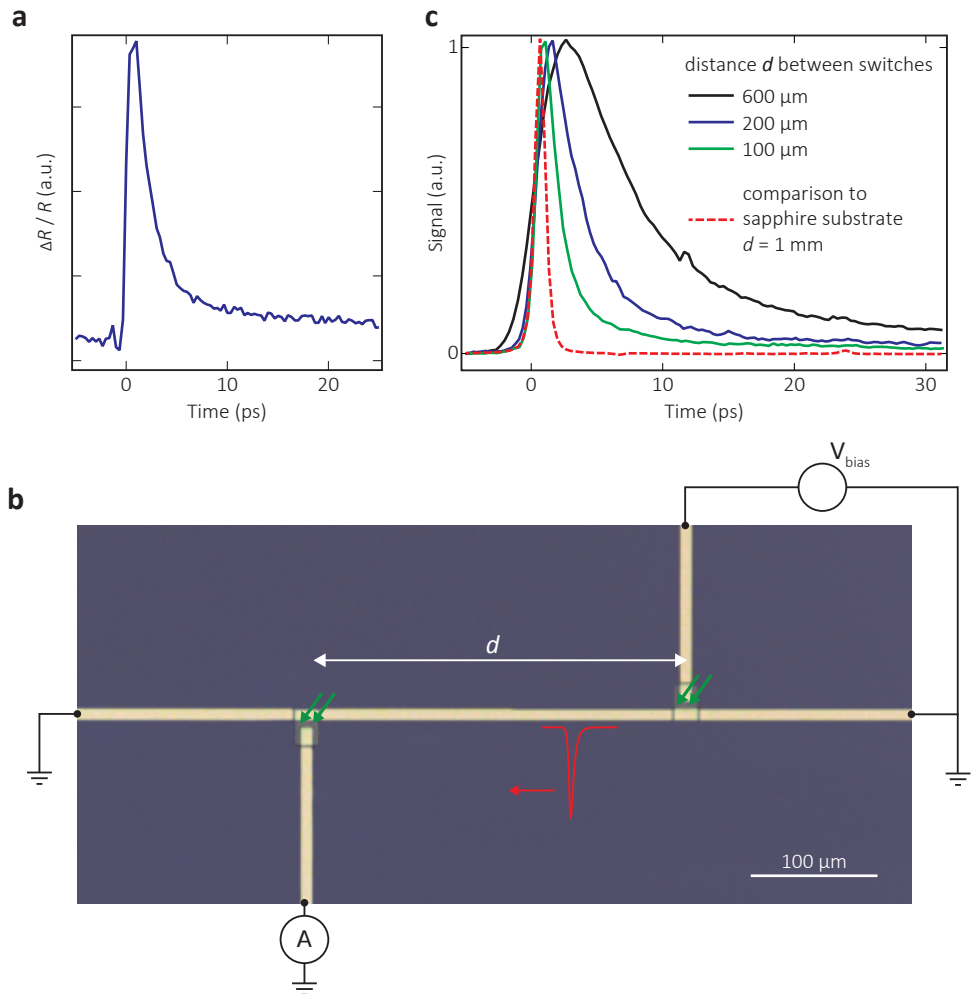

**Fig. S6 | Characterization of the photoconductive switch response time and propagation-induced signal dispersion in the transmission line. a,** Time-resolved reflectivity measurement of an evaporated silicon film used to construct photoconductive switches. The data indicates a carrier recombination time of $\sim$ 1 ps. **b,** Microscope image of a typical calibration circuit consisting of two photoconductive switches connected by a transmission line. When the biased switch (right) was illuminated with a $\sim$ 500 fs laser pulse, a current pulse was launched into the main transmission line. After propagating a distance $d$ down the transmission line, the signal was sampled by a second switch (left). **c,** Sampled current signal at the left photoconductive switch in b, normalized to the peak amplitude. Three calibration circuits built on silicon wafers with $d$ = 100 μm, $d$ = 200 μm, and $d$ = 600 μm shown in b were measured to characterize the propagation-induced signal dispersion in the transmission line. Dashed red line is measured with a calibration circuit built in a coplanar ground-signal-ground waveguide geometry on a sapphire substrate with $d$ = 1 mm, where we find dispersion is negligible on these length scales, to demonstrate the raw photoconductive switch response time.

In this section we characterize the photoconductive switch response time, propagation-induced signal dispersion in the transmission line, and discuss the procedure used to reconstruct the current signals $I_g$ in the main text.

**Photoconductive switch response.** We tested the responsiveness of our photoconductive switch design in two ways. (1) Time-resolved optical reflectivity measurements of the evaporated silicon we used to make photoconductive switches and (2) ultrafast current measurements in a calibration circuit comprised of two photoconductive switches.

*Reflectivity measurements.* We optically characterized the carrier lifetime in our silicon films used to make photoconductive switches by performing time-resolved reflectivity measurements. The result in Fig. S6a shows a carrier recombination time of ∼ 1 ps.

*Calibration circuit measurement.* While the optical measurement shows the carrier lifetime of a clean evaporated silicon film, multiple steps of photolithography are needed to construct photoconductive switches, including the evaporation of Ti / Au contacts. Before performing measurements at low temperatures, the photoconductive switches are laser annealed at room temperature reducing their resistance from hundreds of MΩ to tens of MΩ. This process is known to significantly decrease the carrier recombination time. To characterize the typical photoconductive switch response time we perform ultrafast current measurements in a calibration circuit. The circuit used was similar to that shown in Fig. S6b, but built on a sapphire substrate using a coplanar ground-signal-ground waveguide geometry. Compared to the propagation-induced dispersion observed for microstrips build on Si/SiO$_2$ substrates, we have found through a series of tests that ultrafast signals generated by photoconductive switches exhibit little dispersion when propagating ∼ few hundred microns in a coplanar ground-signal-ground waveguide on sapphire substrates. The dashed red curve in Fig. S6c corresponds to a propagation distance of ∼ 1 mm in such a device and exhibits a nearly perfectly symmetric lineshape, showing that propagation-induced signal dispersion is indeed a small effect over these distances for this transmission line geometry. Therefore, they are well suited to characterizing the raw photoconductive switch response time.

In this measurement, the right switch (Fig. S6b) was biased with a DC voltage and used to launch an ultrafast current pulse into the main transmission line, which was probed by the left switch. By varying the time delay $t - t'$ between the two laser pulses that triggered the switches, the temporal profile of the transient current pulse was obtained. The size of the illuminated area on the switch (< 5 µm) was small compared to the spatial distribution of the transient pulse and is therefore assumed to be a point-like detector. The result for a sapphire device with the two switches separated by 600 µm is shown in Fig. S6c (dashed red line). The signal is symmetric and has a Gaussian width of σ ≈ 425 fs. This signal can be understood as a crosscorrelation of the response functions of the generation switch ($GS$) and detector switch ($DS$). Assuming the generator and detector switches have identical Gaussian-like response functions, the measured Gaussian autocorrelation signal is a factor of $\sqrt{2}$ wider than the response function for an individual switch. We thus derive a response time of σ ≈ 300 fs for our photoconductive switches.

***Propagation-induced signal dispersion in the transmission line.*** The devices used to measure ultra-fast light-induced anomalous Hall currents in graphene were built on doped Si/SiO$_2$ substrates so that the substrate could simultaneously be used as a global backgate to adjust the graphene Fermi level. While ultrafast signals can propagate in microstrip transmission lines built on these substrates, they experience some propagation-induced dispersion. We characterized the propagation-induced signal dispersion in this transmission line geometry by measuring several calibration circuits built on silicon substrates with different distances between the switches. The results are shown in Fig. S5b. As the distance $d$ between the switches was increased, the measured signal broadened, demonstrating that propagation-induced dispersion occurs for circuits built on Si/SiO$_2$.

***Signal Reconstruction.*** To reconstruct the anomalous Hall current signal $I_g$ in the graphene, we first corrected the measured Hall signal $I_x$ (Fig. 2, main text) for propagation-induced signal dispersion to recover the undispersed Hall signal $I'_x$ that was generated at the graphene-microstrip interface ($x$ = 0), then accounted for the RC low-pass effect caused by the capacitance of the graphene flake to obtain $I_g$. To recover $I'_x$, we first needed to determine the effective transfer function $DISP$ describing the propagation-induced signal dispersion, so that it could be deconvolved from $I_x$.

$I_x$ can be described as the convolution of $I'_x$, $DISP$, and $DS$. Note that the detection with a photoconductive switch is described by the crosscorrelation of the signal passing by the switch and $DS$. For a better overview in this section we quote both, dispersion and detection, as a convolution, since it does not change the argumentation. Because $DS$ was fast compared to $I_x$, we assumed that $DS$ could be treated as a delta function so that $I_x$ is given by:

$$I_x \approx I'_x * DISP \tag{6}$$

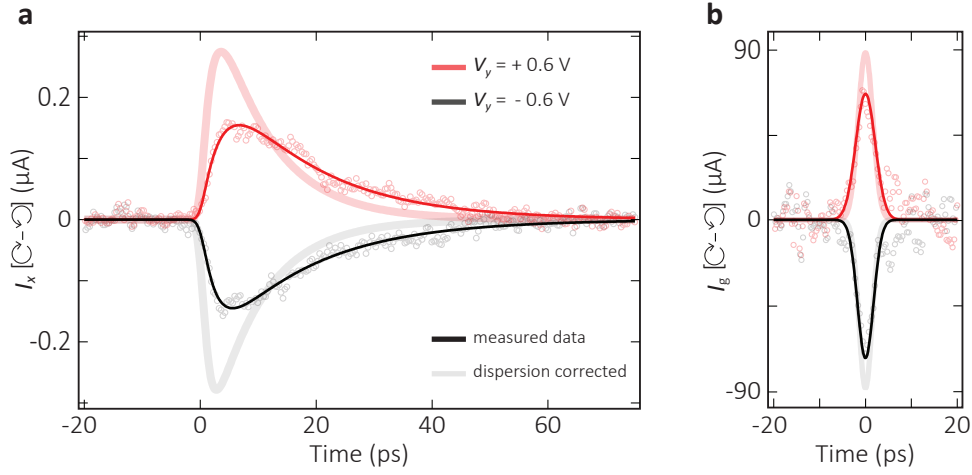

**Fig. S7 | Signal reconstruction. a**, Time-resolved helicity-dependent anomalous Hall currents $I_x[\circlearrowright - \circlearrowleft]$ measured at the photoconductive switch for a positive (red, solid) and negative (black, solid) transverse source-drain voltage $V_y$. The wide light lines show dispersion corrected data. The graphene Fermi level was gated to the Dirac point ($E_F = 0$). **b**, Reconstructed anomalous Hall current signals $I_g$ from the raw measured data (solid) and the dispersion corrected data (light).

We estimated $DISP$ by performing measurements on a calibration circuit built on a silicon substrate with the switches 200 µm apart (Fig. S6c, blue curve), where we expect the propagation induced signal dispersion to be approximately equal to that in the graphene device. The measured signal comprises of three elements: The response function of the generation switch ($GS$), the dispersion induced by the transmission line ($DISP$), and the response function of the detection switch ($DS$). The signal is thus the convolution $GS * DISP * DS$. Since $GS$ and $DS$ are fast compared to the measured signal, they can be approximated as delta functions, and the convolution becomes $GS * DISP * DS \approx DISP$. Thus, by considering conservation of charge and deconvolving $I_x$ with the blue curve in Fig. S6c, we obtained $I'_x$ (Fig. S7a). The data show that the dispersion corrected signal $I'_x$ becomes shorter and the amplitude increases in comparison to $I_x$.

As discussed on p. 9, the signal $I'_x$ that is discharged into the transmission line from the graphene at $x = 0$ is related to the anomalous Hall currents $I_g$ generated in the graphene via the transfer function for a low-pass filter. Thus, we deconvolved the dispersion corrected $I'_x$ data in Fig. S7a with the transfer function for a first-order low-pass filter, which we used to model the current generation (see Eq. 3). We determined $\tau_{\text{RC}}$ by fitting the exponential decay of $I'_x$. Figure S7b (light lines) shows the reconstructed current signals accounting for dispersion, which show a time duration of $\sim 3$ ps (FWHM). The increased noise in the reconstructed data in Fig. S7b is caused by the deconvolution process. The peak amplitude of the Hall signal increases by a factor of $\sim 6.0$ when comparing the reconstructed signal with the originally measured trace. When dispersion is not accounted for in this analysis, the reconstructed signals have a time duration of $\sim 4.5$ ps (FWHM).

## S3.3: Calibrating the photoconductive switch

In the following we will lay out the procedure used to calibrate a photoconductive switch in conjunction with our microstrip transmission line geometry. We present two different ways to calculate the net charge contained in a pulse traveling in a microstrip with a specific geometry. Both calibration methods deliver similar values for the charge. Thus, the procedure delivers a quantitatively correct measure of the charge transmitted through the microstrip. First we demonstrate that one can determine the amount of charge per pulse generated by a photoconductive switch. With this being established we scale a remotely measured time-resolved signal of the generated pulse to reconstruct the current profile. The second approach directly relates a known voltage at the position of the photoconductive switch to a measured signal. By characterizing this relationship with a DC voltage bias we can quantitatively determine the voltage profile of the pulse traveling down the microstrip.

The microstrip transmission line used here propagates only the fundamental quasi-TEM mode in the relevant frequency range (up to 2 THz). This is a key reason why we can make quantitative conductance measurements using this transmission line geometry. Specifically, we use microstrips build on $SiO_2$ / $Si^{++}$ wafers to propagate pulses. The silicon wafer (Nova Electronic Materials, 6'' P <100>, prime grade) consists of 700 µm p-doped silicon (0.001-0.005 Ωcm) and a top layer of 285 nm $SiO_2$. The microstrip is chosen to be 5 µm to 20 µm wide and ~ 300 nm thick, depending on the size of the graphene flake being studied. With these dimensions, the fundamental mode of the microstrip as depicted in Fig. S8a is the only excited mode[38]. In principle higher order quasi-$TEM_{mn}$ modes could be excited. However, since the distance between backplane (doped Si) and microstrip as well as the width of the microstrip are both much smaller than the guided signal wavelength (0.3 µm and 6 µm << 100 µm - 300 µm, see Fig. S8b) those cannot be coupled and propagated in our microstrip design. Such modes have been studied in detail by applying full wave analysis methods to the microstrip geometry confirming the above findings explicitly for the sub-picosecond regime, e.g. in Ref. 39. The first higher order mode relevant for the chosen design is the quasi-$TE_{10}$ whose cut-off frequency $f_{c,TE10}$ can be evaluated by setting m = 1 in the expression

$$f_{c,\text{TEm0}} = m \frac{c}{2w_e\sqrt{\varepsilon_r}},$$

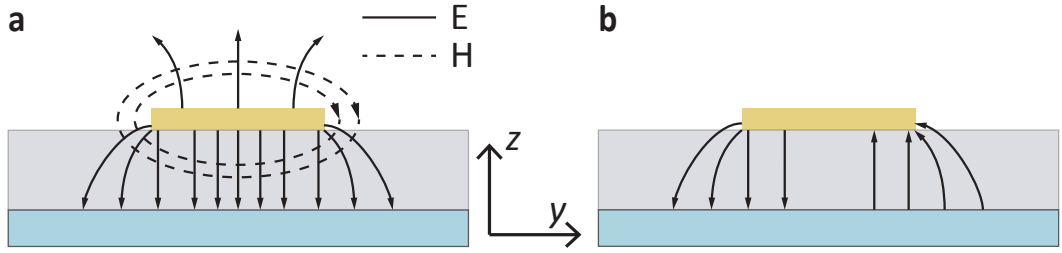

**Fig. S8 | Electric field distribution for microstrip modes. a,** Fundamental mode. **b,** First higher order quasi-TE$_{10}$ mode, which cannot be excited in this experiment, where the distance between backplane (blue) and the microstrip as well as the width of the microstrip are both much smaller than the guided signal wavelength (0.3 µm and 6 µm << 100 µm - 300 µm ).

where $c$ is the speed of light, $w_e$ is the effective strip width and $\varepsilon_r$ is the frequency dependent dielectric constant of SiO$_2$. Thus the cut-off frequency for the first higher order mode in the present microstrip is in the ten THz regime, clearly above the frequency of the guided signals we observe.

For later analysis, we determine the impedance of the microstrip using the transmission line calculator in QucsStudio[40]. We use a dielectric constant $\varepsilon_r$ = 3.9 for SiO$_2$, a resistivity of $\rho_{gold}$ = 2.4 · 10$^{-8}$ Ωm at a frequency of 1 THz. We find that the results are in good agreement with full wave analysis finite elements simulations. The microstrip geometry ($w$ = 5.6 µm) used in the experiment discussed in the main text yields an impedance $Z_0$ = 8.6 Ω. For the measurements performed to verify the calibration schemes discussed in the following, we used test samples with a width $w$ = 10 µm, which yields an impedance $Z_0$ = 5.1 Ω. We find that for our microstrip design current and voltage are in phase and related by the impedance $Z_0$.

***Lock-in amplifier preliminaries.*** Signals in the experiment are detected using a Stanford Research System lock-in amplifier SR830. We will briefly summarize the specifics of this lock-in amplifier (LIA). Please note that the following can be different for different LIA models depending on the internal reference wave form and internal electronics. All values reported are amplitudes or peak to peak values, while the SR830 reports rms values of the Fourier components[41]. Therefore, the sine wave input with amplitude $V_{\text{IN,sine}}$ and the LIA reading $V_{\text{LIA}}$ are related by:

$$V_{\text{IN,sine}} = \sqrt{2}\, V_{\text{LIA}}.$$

To obtain quantitative measurements it has to be considered that the LIA measures the Fourier components of the input signal. This is important when comparing square and sine wave forms. The Fourier series of a square wave with amplitude of 1 V (2 V peak to peak) is given by:

$$f(t) = 1.237 \sin(\omega t) + 0.4244 \sin(3\omega t) + 0.2546 \sin(5\omega t) + \cdots ,$$

where the angular frequency is given by $\omega = 2\pi f$. Thus, the amplitude of the square wave input with amplitude $V_{IN,square}$ and frequency $f$ are connected to the LIA reading $V_{LIA}$ in the following way:

$$V_{IN,square} = \frac{\sqrt{2}}{1.237} V_{LIA}.$$

For example when we send a square wave $V_{IN,square}$ = 1 V into the LIA, the LIA will report $V_{LIA}$ = 0.87 V.

***Determination of the net charge in a propagating pulse.*** In this section, we demonstrate that one can determine the net charge contained in a pulse traveling in a microstrip. If the microstrip is designed to only propagate a quasi-TEM mode, the total charge in one pulse is an important element to reconstruct the current profile.

A schematic of the test sample is shown in Fig. S9a. Two photoconductive switches are connected to a central microstrip. First, we investigate the behavior of the left switch that we use as a pulse generator in this part of the discussion. In general the pulse could also originate from a different source with a sufficiently high DC resistance, for example a graphene flake. The second photoconductive switch on the right side will be used later to characterize the temporal profile of the generated pulse, but can be neglected for the following discussion.

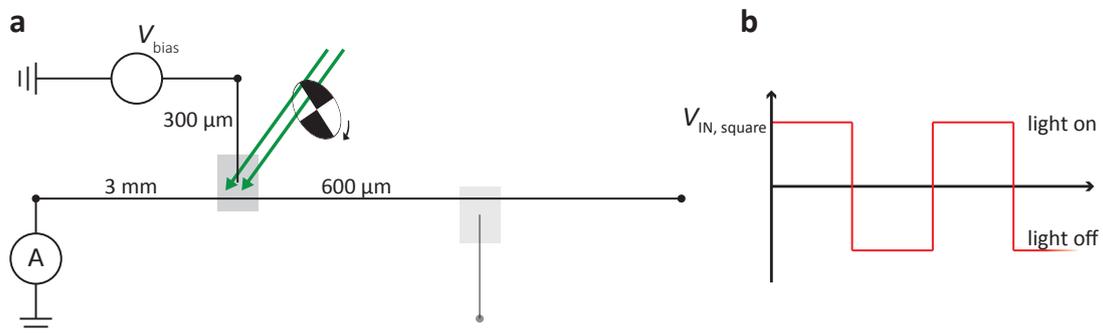

**Fig. S9 | Characterization of a photoconductive switch. a,** Schematic representation of the experimental setup used to characterize a photoconductive switch. The switch is held under a DC voltage bias ($V_{bias}$) and the average current generated upon illumination with a train of laser pulses ($f_{rep}$ = 211 kHz) is measured at the chopping frequency of 409 Hz by a TIA and LIA (represented by the ammeter). **b,** The light induced current is given by the peak-to-peak signal detected by the LIA.

To measure the charge injected by the photoconductive switch, the switch is biased with a known voltage and we measure the generated current upon illumination with a pulsed laser source with a repetition rate $f_{rep}$ = 211 kHz. The generated signal is amplified by a home-built transimpedance amplifier (TIA) with a bandwidth $BW$ = 8 kHz and an amplification $A_{TIA}$ = 2·10$^9$ V/A. DC input signals are blocked with a 1 µF. The high pass cut-off frequency depends on the resistance of the photoconductive switch and is ~ 1 mHz. Measurements are performed at 80 K where the DC switch resistance is ~ 150 MΩ. At the frequency of 407 Hz the input impedance of the TIA is 390 Ω and is several orders of magnitude smaller than the switch resistance. Thus, the whole bias voltage drops across the photoconductive switch and the input of the TIA has practically ground potential. The amplified signal is recorded using a LIA. The laser light is periodically turned on and off at a frequency of 407 Hz by a rotating chopper blade. Thus, the modulation of the laser intensity and the generated current follow a square signal as depicted in Fig. S9b and we need to consider the above mentioned correction for square waves. Figure S9b shows that the light induced current is given by the peak to peak value of the square wave. Therefore, the LIA output $V_{LIA}$ and the average current $I_{light}$ are connected by

$$I_{light} = \frac{2\sqrt{2}}{1.237} \frac{1}{A_{TIA}} \cdot V_{LIA} . \qquad (7)$$

Figure S10 shows the generated signal as a function of bias voltage. From this we calculate the relation between applied bias voltage $V_{bias}$ and the measured signal $V_{IN}$, which is proportional to the generated current:

$$c_1 = \frac{V_{IN}}{V_{bias}} = \frac{\sqrt{2}}{1.237} \frac{V_{LIA}}{V_{bias}} = 0.8 \pm 0.05 . \qquad (8)$$

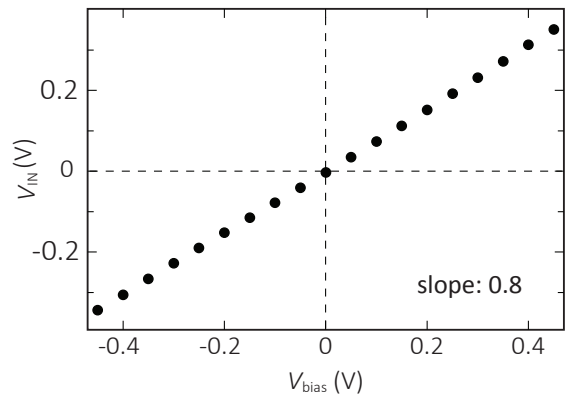

**Fig. S10 | Measured signal as a function of applied bias voltage.** The measured signal is proportional to the average current flowing and therefore proportional to the amount of charge generated per laser pulse triggering the switch.

The error of the calibration factor (Eq. 1) arises from a small nonlinearity of the switch. When the switch is illuminated by a laser pulse, a picosecond current pulse is generated. This pulse will stretch in the microstrip and reflect at the contacts, but in the end the net charge of the pulse will leave the sample when measuring on long time scales (407 Hz). It is important to understand that this statement only implies that we are able to measure the ~ DC (407 Hz) component of the generated pulses in our specific microstrip geometry, which is equal to the integral of the pulse. This can be critical when dealing with bipolar pulses.

To demonstrate that the switch acts like a current source and no charge is lost we insert a 10 kΩ resistor in front of the TIA (see Fig. S11a). In Fig. S11b the result of this measurement is compared to the measurement without a resistor. We find no difference. This is explained by the high resistance of the switch (~ 150 MΩ), which is equivalent to a current source (measured at 407 Hz) with a very high internal resistance compared to the load. Furthermore, this shows that the length of the microstrip has no impact on this kind of measurement.

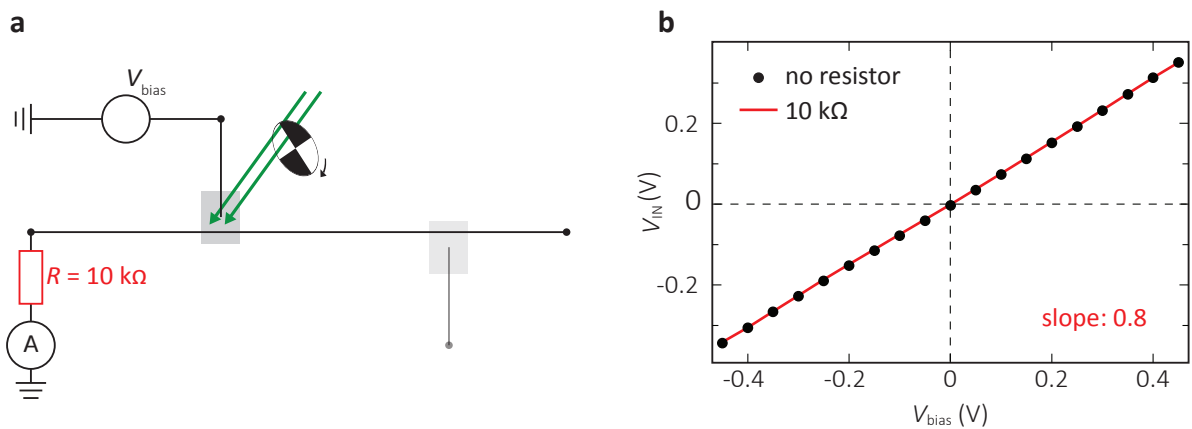

**Fig. S11 | Termination with a 10kΩ resistor. a,** Schematic representation of the experimental setup. A 10 kΩ resistor is inserted in front of the TIA. **b,** The measured signal is unchanged when the resistor is inserted. This suggests that, on long timescales, a photoconductive switch acts like an effective current source with a high internal resistance and all of the generated charge can be measured.

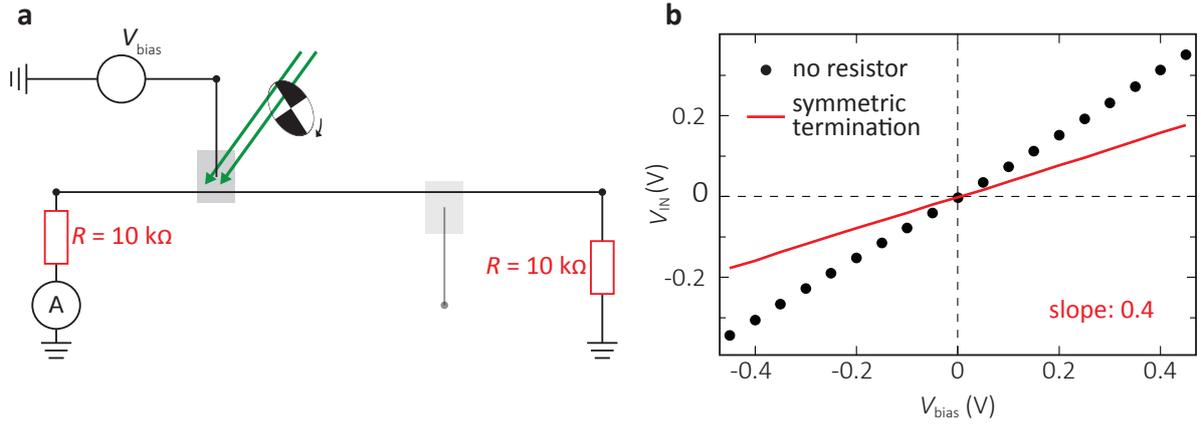

**Fig. S12 | Symmetric termination of the central microstrip. a,** Schematic representation of the experimental setup with a symmetric 10 kΩ termination. **b,** Measured signal compared to a measurement with a termination on one side only.

Next, we ground both sides of the central microstrip and measure the current on the left side (see Fig S12). The result is shown on the right side in Fig. S12. As expected the current can flow through both resistors and therefore the current measured on one side is halved.

We perform these control procedures to demonstrate that the time integrating measurement is a robust way to determine the average current upon laser illumination and from there the charge flowing per laser pulse:

$$Q_{\text{light}} = \frac{I_{\text{light}}}{f_{\text{rep}}} = \frac{2\sqrt{2}}{1.237} \frac{1}{A_{\text{TIA}} f_{\text{rep}}} \cdot V_{\text{LIA}}. \tag{9}$$

Of course, this does not teach us anything about the actual current profile in the microstrip since we are missing the time resolution. Also, in the case of an oscillatory signal, which on average does not transfer charge, this kind of measurement would be meaningless. But in the case of the simple unipolar pulse that we want to characterize, the measured charge can be used to reconstruct the current profile at the position of a second photoconductive switch as it is discussed in the next section.

***Switch calibration.*** In this section we use two different schemes to reconstruct the quantitative current or voltage profile of a pulse traveling in the microstrip.

Figure S13 depicts the experimental setup. Two switches are connected to a central microstrip. The right switch (switch 2) is biased with $V_{bias}$ = 5 V and is used as a pulse generator. Using this switch as a test source for short pulses, we will characterize the pulses traveling down the microstrip. The left switch (switch 1) is 600 µm away and used to detect the generated pulses. The laser light on the generator switch is intensity chopped at a frequency of 407 Hz. Thus, the LIA behind the detector switch will only detect changes that are induced by the pulses passing by. By varying the time delay between the pulses triggering the detector and generator, the temporal profile of the pulse is recorded. The result is shown in Fig S14. The photoconductive switch probes the electric field lines that bias the switch. For the fundamental quasi-TEM mode propagating in the microstrip (see Fig. S8a), the detected signal is proportional to the current in the microstrip as current and voltage are in phase and related by the impedance $Z_0$. The temporal profile of the pulse represents the pulse after it propagated 600 µm in the microstrip. In the following discussion we will quantitatively characterize the pulse at the position of switch 1.

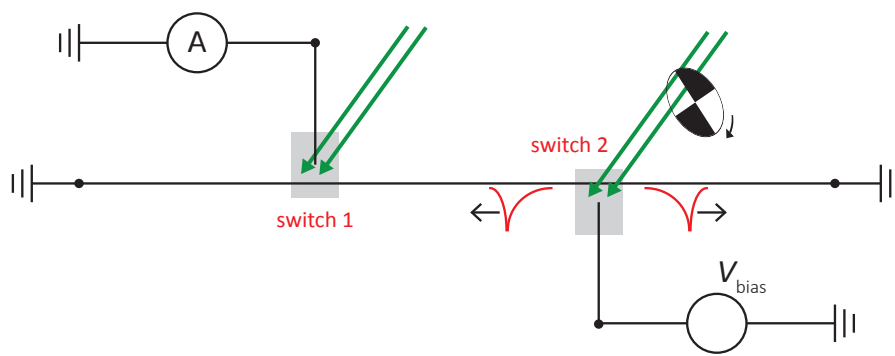

**Fig. S13 | Pump-probe measurement setup.** Switch 2 (right) is held under a DC voltage bias and used as a generator for short pulses. By varying the time delay between triggering the detector switch 1 (left) and switch 2, the temporal profile of the pulse is recorded.

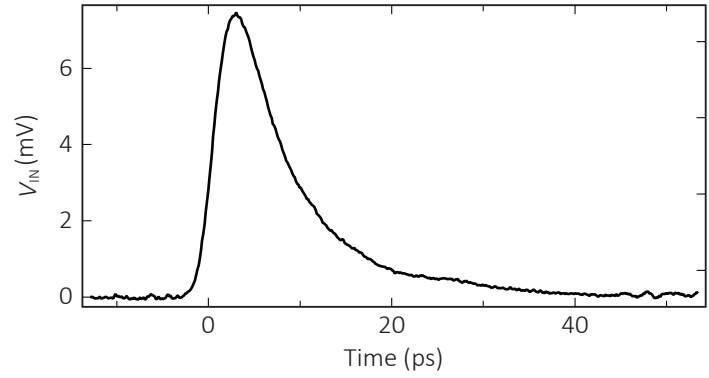

**Fig. S14 | Pump-probe signal.** The generator switch is biased with $V_{bias}$ = 5 V and sampled by a second switch 600 µm away. The curve is scaled to report the amplitude of the input signal into the LIA after being amplified with $A_{TIA}$ = 2·10$^9$ V/A.

***Method 1.*** First, we will exploit the fact that we can measure the total amount of charge $Q_{light}$ injected into the microstrip per laser pulse hitting the generator switch (Eq. 9). When $Q_{light}$ is injected into the microstrip, two identical pulses are generated[*] (Fig. S13) that propagate in opposite directions, each containing charge:

$$Q = \frac{Q_{light}}{2} = \frac{\sqrt{2}}{1.237} \frac{1}{A_{TIA} \, f_{rep}} \cdot V_{LIA} \, .$$

Here $V_{LIA}$ is measured at the end of the central microstrip with only one side grounded (Fig. S9a, with $V_{bias}$ and the laser light on the right switch). The integral of a single unipolar pulse sampled by switch 1 (Fig. S13) is thus related to the charge Q within that pulse, which is what we will determine in the following.

With an applied bias voltage $V_{bias}$ = 5 V across switch 2 we measure $V_{LIA}$ = (5.9±0.2) V, which yields the amount of charge carried in one pulse:

$$Q = (1.7 \pm 0.1) \cdot 10^{-14} \, C$$

---

[*] The previous discussion about measuring the average current generated by a photoconductive switch, which behaves like a current source, is only valid when measuring at low frequencies (in other words on long time scales). When performing on-chip measurements, the pulse is sampled in real time. On short time scales, the pulse travels a small distance and perceives the local impedance of the microstrip. Thus, the pulse cannot distinguish between a microstrip that is grounded or open at the end. As a consequence, when switch 2 is hit by a laser pulse, two pulses are launched into the central microstrip. One pulse travels to the left and one pulse travels to the right (see Fig. S13), independent of whether or not the central microstrip is grounded on one or two sides. However, at least one side of the central microstrip needs to be grounded, otherwise only a fraction of the bias voltage Vbias would droop over switch 2 and the measurement of the amount of charge per pulse at the end of the central microstrip (compare to Fig. S9) and the time resolved measurement (Fig. S13) would not be equivalent.

As discussed previously, the signal measured with a photoconductive switch is proportional to the current profile of the pulse traveling in the microstrip. Thus, the signal shown in Fig. S14 represents the current profile of the investigated pulse and the integral of this curve is equal to the amount of charge $Q$. This assumes that only a quasi-TEM mode propagates in the microstrip and that the voltage and current profile are in phase. This holds true for the simple microstrip geometry used in the experiment. The resulting current profile is shown as the blue curve in Fig. S15a.

***Method 2.*** For the second approach to quantitatively characterize the current pulse at the position of switch 1 we determine the response of the detector to a known bias voltage. We then reconstruct the voltage profile and from there the current profile. A known voltage $V_{bias}$ is applied across switch 1 and we measure the current at the end of the microstrip, analogue to the calibration in Fig. S9. From Eq. 8 we know the relation between the measured signal $V_{IN}$ (Fig. S10) and the bias voltage $V_{bias}$ necessary to generate this signal. If we suppose that the transient pulse acts like a voltage biasing the switch, we can obtain the voltage and current profile:

$$V_{bias} \cong V_{pulse} = \frac{V_{IN}}{c_1}, \qquad I_{pulse} = \frac{V_{pulse}}{Z_0}.$$

The resulting current profile is shown as the red curve in Fig. S15a and is in good agreement with the current profile obtained from method 1.

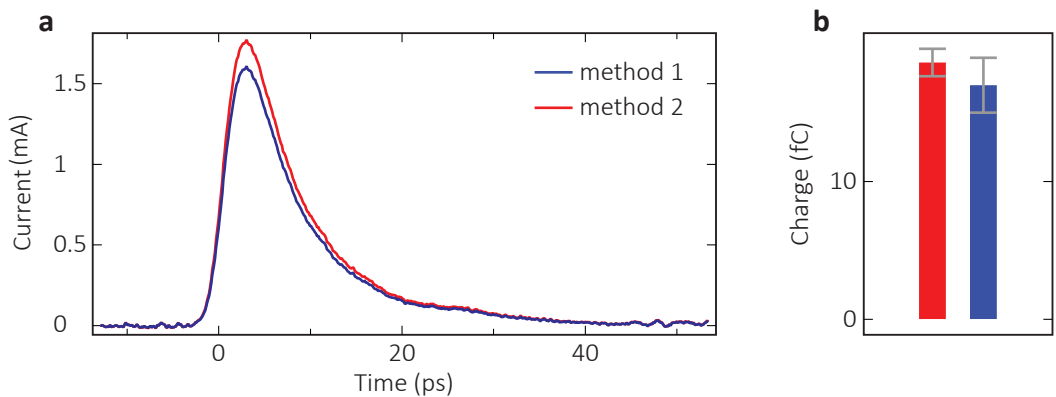

**Fig. S15 | Comparison of the two methods to quantitatively characterize the transient current in the microstrip. a,** Reconstructed current profile according to method 1 (blue) and method 2 (red). **b,** Total amount of charge contained in one pulse obtained from the two different methods to characterize the pulse. The two methods are equivalent within the error bars.

What this shows is that the charge and electric field distribution of the transient pulse and the DC voltage bias are comparable, as we expect it to be for a quasi-TEM mode and signals longer than the switch decay time. The comparison of the charge obtained from the two methods shown in Fig. S15b proves this to be true within the error bars.

The above analysis verifies that quantitative measurements of the charge flowing in a microstrip transmission line can be made on ultrafast timescales using a photoconductive switch. We discuss how this analysis is applied to extracting the anomalous Hall conductance in graphene in the following section.

## S3.4: Estimation of the non-equilibrium anomalous Hall conductance

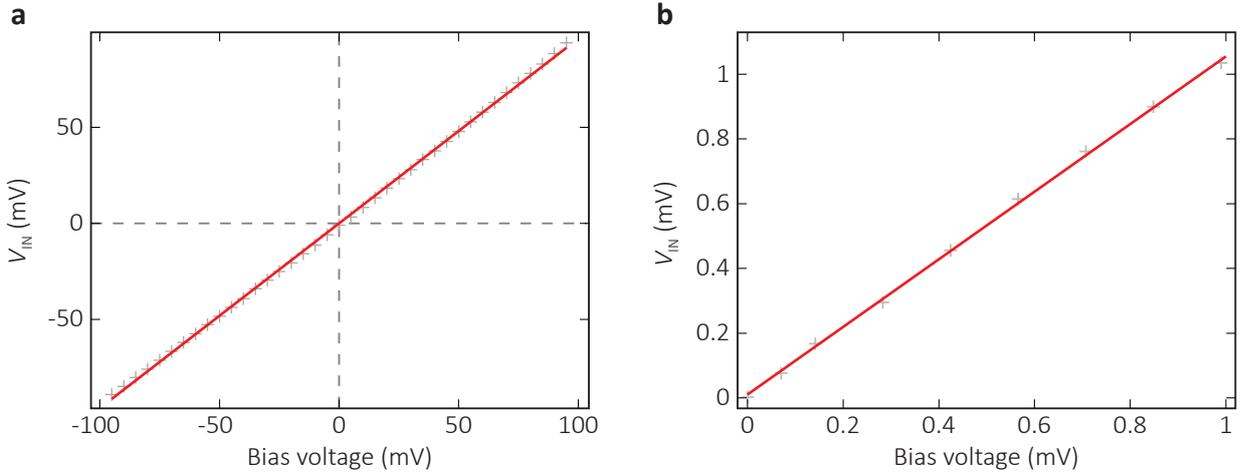

**Fig. S16 | Circuitry calibration.** Measured lock-in amplifier signal of a DC-voltage biased switch triggered with an intensity chopped laser beam. **a**, Calibration for bias voltages in the 100 mV range. **b**, Calibration for bias voltages in the 1 mV range.

The current signal $I_x(x, t)$ propagating in the transmission line carries a voltage $V_x(x, t)$, where the two are related via the transmission line impedance $Z_0 = V_x(t, x) / I_x(t, x) \sim 8.6\pm1.0\ \Omega$. At the switch position, we measure a signal using a lock-in amplifier (LIA) that is directly proportional to $V_x(t, x_s)$. To quantitatively determine $V_x(t, x_s)$, we needed to calibrate the LIA reading to a known voltage bias as discussed in the previous chapter.

To do this, we biased the switch in the graphene device used in the main text with a known DC-voltage and detected the LIA signal upon laser illumination (*Method 2* in "switch calibration" of the previous section). The graphs in Fig. S7 show the measured LIA signal for a given voltage biasing the switch. A calibration factor $V_{IN}/V_{bias}$ of $c_1 = 0.96$ and $c_2 = 1.05$ were determined for 100 mV and 1 mV measurement ranges, respectively. The LIA signals in the Hall current experiment were tens of μV. Thus, we used a calibration factor of $c_2 = 1.05\pm0.10$ for the analysis.

The curves shown in Fig. 2a and Fig. 3 (main text) are calibrated taking into account the sine wave modulation introduced by the polarization chopping (see p. 20), a calibration factor $c_2 = 1.05$, and an impedance of $Z_0 = 8.6\ \Omega$. The plotted current $I_x[\circlearrowright - \circlearrowleft]$ is given by the peak-to-peak signal of the LIA output $V_{LIA}$ (compare to the light induced current discussed in Fig. S9b):

$$I_x[\circlearrowright - \circlearrowleft] = \frac{V_x[\circlearrowright - \circlearrowleft]}{Z_0} = \frac{2 \cdot V_{IN}}{Z_0 c_2} = \frac{2\sqrt{2} \cdot V_{LIA}}{Z_0 c_2}$$

In the general case, the transverse current is given by $I_x = G_{xx}V_x + G_{xy}V_y$. On ultrafast timescales, Hall currents generated in the graphene perceive the low input impedance $Z_0$ of the transmission line and leave the sample. Thus, in contrast to DC Hall effect measurements, there is no $V_x$ Hall voltage accumulation in the graphene and the transverse current is described by $I_x = G_{xy}V_y$. The peak non-equilibrium anomalous Hall conductance is thus given by $G_{xy} = \hat{I}_g[\circlearrowright - \circlearrowleft] / 2V_y$, where $\hat{I}_g$ is the peak of the reconstructed signal in Fig. 3b (main text). The slope of the fitted line in Fig. 3a (main text) gives a value of $\hat{I}_x[\circlearrowright - \circlearrowleft] / V_y \approx (2.5 \pm 0.5) \cdot 10^{-5}\,\Omega^{-1}$. The reconstruction of $\hat{I}_g$ increases this ratio by a factor of $\sim 6.0$, which yields a peak Hall conductance:

$$G_{xy} = \frac{\hat{I}_g[\circlearrowright - \circlearrowleft]}{2V_y} = \frac{6.0}{2} \cdot \frac{\hat{I}_x[\circlearrowright - \circlearrowleft]}{V_y} \approx 1.8 \pm 0.4 \,\frac{e^2}{h}.$$

The error of the peak Hall conductance arises from small nonlinearities of the switch response (see Fig. S7), the difference between the two calibration methods discussed in section 3.3, which is represented by the uncertainty of the transmission line impedance $Z_0$, and the statistical error.

# S4. Photocurrent background at zero source-drain voltage

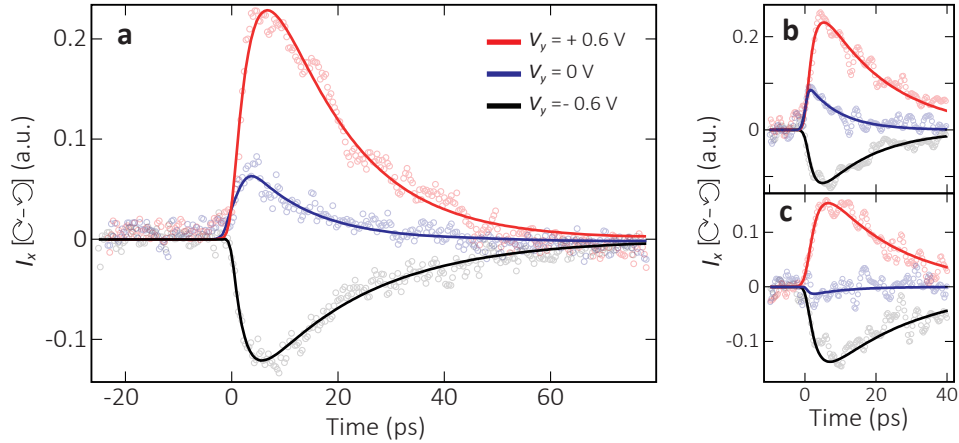

**Fig. S17 | Raw Hall current data including a zero source-drain voltage background signal.**
**a**, $I_x[\circlearrowright - \circlearrowleft]$ data from Fig. 3a (main text) including the zero source-drain voltage background signal. **b**, and **c**, $I_x[\circlearrowright - \circlearrowleft]$ taken after two different thermal cycles.

A small helicity-dependent photocurrent signal was sometimes observed with no source-drain voltage applied. This signal was recorded during every measurement run and subtracted as a background. We note that after background subtraction, the signals for positive and negative $V_y$ were always symmetric. This background signal likely results from an effective extra bias voltage caused by band bending fields at the contacts.

Figure S17a shows the $I_x[\circlearrowright - \circlearrowleft]$ data presented in Fig. 2 (main text) including the background signal at $V_y = 0$. This signal exhibited no measurable dependence on the angle of incidence of the light, which rules out the circular photon drag effect[42]. It could be a shift current[43] or a circular photogalvanic current[42,] generated at the edge of the graphene flake. While this is an exciting prospect, we believe the signal is more likely due to inhomogeneous band-bending fields at the probing contact. The inhomogeneity can arise as a result of the inhomogeneous doping (responsible for electron-hole puddles) that occurs in graphene[44]. We find that thermal cycling can have a large effect on the relative amplitude of this photocurrent (Fig. S17b-c), in some cases reducing it almost to zero. Based on our observation that the Dirac voltage changes slightly after thermal cycling, we believe that the electron-hole puddle distribution may also be varying. This would in turn affect the inhomogeneity of the band-bending fields. This is an indication that inhomogeneous band-bending fields at the probing contact are the most likely source of this photocurrent.

# S5. Results using a different device geometry

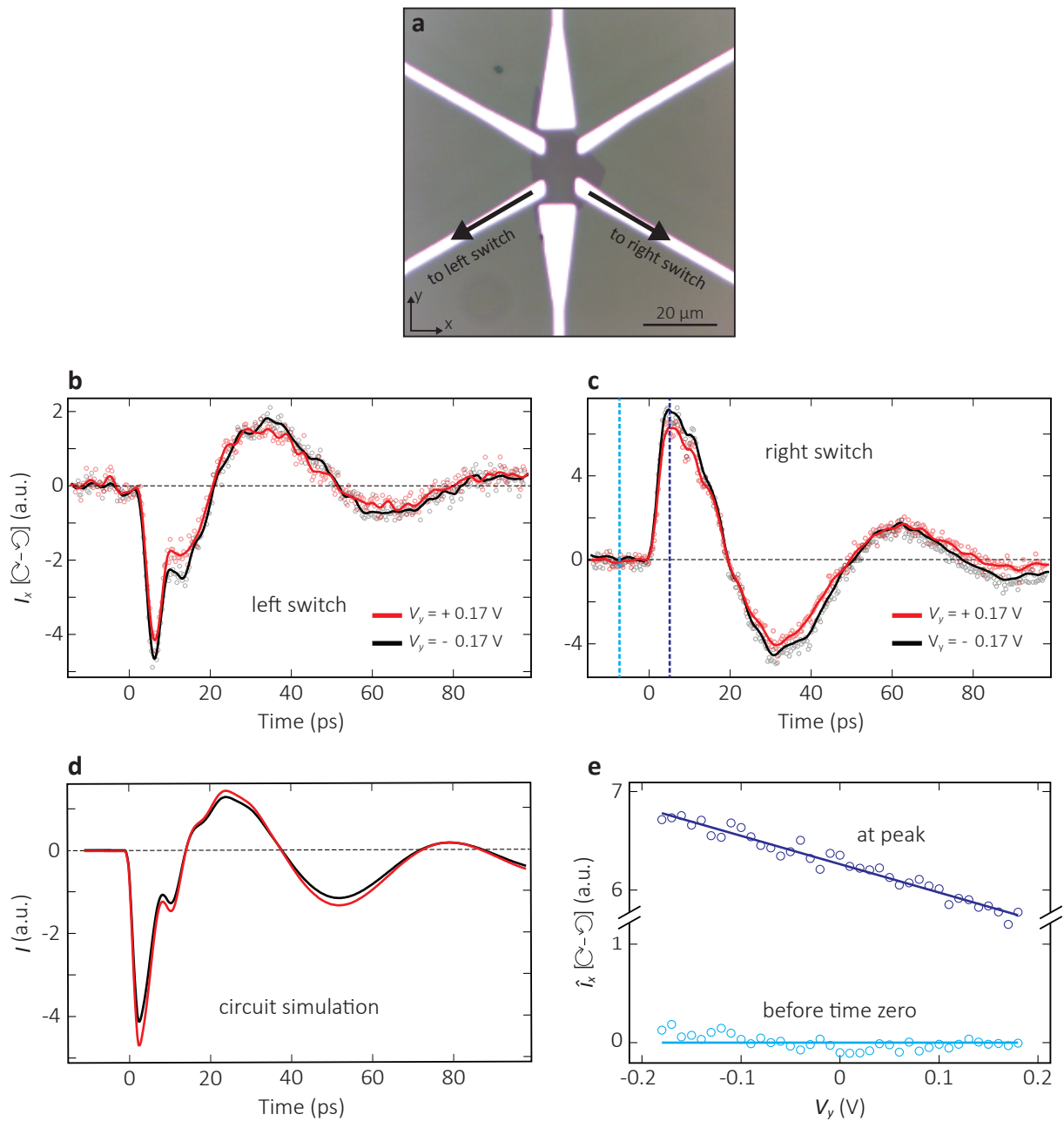

**Fig. S18 | Results from a second device. a**, Optical microscopy image of a graphene device with a different transmission line and contact geometry. Helicity dependent currents $I_x[\circlearrowright - \circlearrowleft]$ measured at the left switch **b**, and right switch **c**, for positive (red) and negative (black) source-drain voltage $V_y$. **d**, Circuit simulation for the device geometry shown in (a). The simulation takes into account the flake capacitance and the transmission line geometry. The circuit is triggered by a picosecond unipolar pulse with varying amplitude. **e**, Helicity dependent currents $\hat{I}_x[\circlearrowright - \circlearrowleft]$ measured as a function of source-drain voltage. The signal is measured before time zero (light blue) and at the peak of the signal (dark blue). The graphene Fermi level was gated to the Dirac point ($E_F = 0$).

Figure S18 shows the results from a device with a different transmission line and contact geometry from that studied in the main text. This device geometry has certain disadvantages, but we have used it to observe that anomalous Hall signals are generated with opposite polarity at opposite sides of the sample, as expected for any Hall effect.

An optical microscopy image of the device is shown in Fig. S18a. The two large electrodes were used to apply a source-drain voltage $V_y$. Two of the side transmission lines contacting the flake were equipped with photoconductive switches. Helicity-dependent currents $I_x[\circlearrowright - \circlearrowleft]$ were measured at the right and left photoconductive switches for positive (red) and negative (black) source-drain voltage $V_y$ (Fig. S18b-c).

The first feature of note in these data sets is that there is a larger signal generated at $V_y = 0$ compared to the device studied in the main text (see Fig. S17 for comparison). This signal is not explicitly plotted in Fig. S18b-c, but it resides in between the signals generated with positive and negative $V_y$. This larger $V_y = 0$ signal is probably due to the asymmetric contact design for this device. For a given probing contact, the band-bending fields in the lower region of the contact can be expected to be modified by the close proximity of the lower source-drain contact. This asymmetry provides a larger effective built-in $V_y$ in the vicinity of the probing contacts compared to the symmetric device design studied in the main text. The gap between the probing contacts and the lower source-drain contact is also smaller than the wavelength of the light used to optically excite the device. Thus, diffraction is expected to reduce the illumination of the graphene in this region. This acts as an additional asymmetry that could contribute to the signal at $V_y = 0$.

The second feature of note is that the signals exhibit large oscillations. These are due to signal reflections at the end of the transmission line, which were $\sim$ 6x shorter compared to those used in the device studied in the main text. The ringing behavior is well modeled by a lumped circuit simulation (see Fig. S18d), similar to the simulations discussed in detail in section S3.1, which takes into account the geometry of this transmission line design. Nevertheless, using this device we were able to observe the emergence of anomalous Hall currents in graphene (Fig. S18e). We also verified that signals of opposite polarity are generated at opposite sides of the sample, as can be seen by comparing Fig. S18b and S18c. Anomalous Hall current signals were also observed on three other devices that explored different device geometries.

# S6. Substrate effects

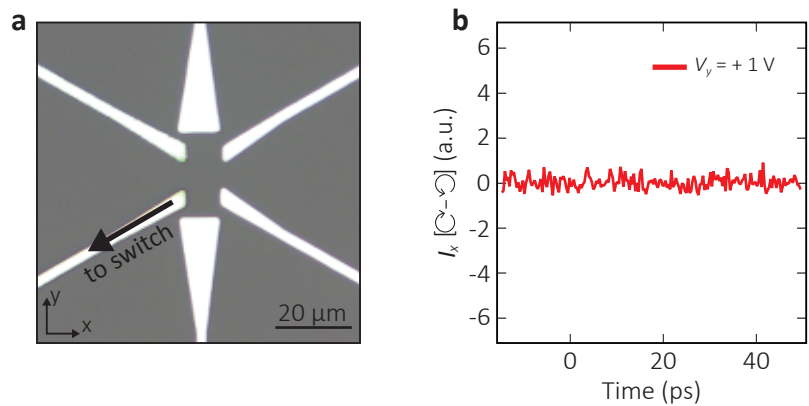

**Fig. S19 | Test measurements on a device without a graphene flake. a**, Optical microscopy image of a device without a graphene flake. The device has the same geometry discussed in section S5. **b,** Helicity dependent currents $I_x[\circlearrowright - \circlearrowleft]$ measured at the left switch.

To evaluate signals that might be created by illuminating the substrate and transmission lines with mid-infrared laser pulses, we performed time-resolved measurements on a sample without a graphene flake. Figure S19a shows the device, which features the same microstrip design used for the device discussed in the previous section. Figure S19b confirms that no signal is generated in this device, verifying that the measured signals associated with the light-induced anomalous Hall effect in the main text originate purely from the graphene.

We note that we do detect a very small signal when an intensity chopped green laser beam shines on the substrate close to the microstrip, which is likely due to the creation of electron-hole pairs in the substrate whose radiative effects are captured by the microstrip. However, this has no bearing on our experimental results performed with circularly polarized mid-infrared laser pulses, where these effects do not appear. This is likely in part because of the low photon energy (∼ 190 meV) of the mid-infrared light compared to the silicon substrate band gap.

## S7. Floquet band structure calculations

The system is modeled by a tight-binding Hamiltonian of the form

$$\widehat{H}_{\text{TB}} = t \sum_{j=1}^{3} \cos(\mathbf{k} \cdot \mathbf{d}_j) \hat{\sigma}_x + \sin(\mathbf{k} \cdot \mathbf{d}_j) \hat{\sigma}_y \qquad (10)$$

where $t$ denotes the tunneling energy, $\hat{\sigma}_{x,y}$ are the standard Pauli matrices in the basis of Bloch functions residing on the two triangular sub-lattices forming the honeycomb lattice, $\hbar \mathbf{k} = \{p_x, p_y\}$ is quasi-momentum and $\mathbf{d}_1 = a\{1/2, \sqrt{3}/2\}, \mathbf{d}_2 = a\{1/2, -\sqrt{3}/2\}, \mathbf{d}_3 = a\{-1,0\}$ are the vectors connecting nearest-neighbour carbon atoms with spacing $a = 1.42$ Å. We set $t \approx 3.09$ eV such that the Fermi velocity $v_F = 3ta/2\hbar = 10^6 m/s$ [*]. The Dirac points, where the two bands of this Hamiltonian become degenerate, appear at momenta $\mathbf{p}_{K,K'} = \{\frac{2\pi}{3a}, \pm \frac{2\pi}{3\sqrt{3}a}\}\hbar$.

In addition, the Hamiltonian in the vicinity of the Dirac points is compared to the approximate Dirac Hamiltonian $\widehat{H}_D = v_F(\tilde{p}_x \hat{\sigma}_x + \tilde{p}_y \hat{\sigma}_y)$ with $\tilde{\mathbf{p}} = \mathbf{p} - \mathbf{p}_K$. Within the range of energies explored in Fig. 4 (±0.2 eV, main text) the static eigenenergies of the two Hamiltonians differ by less than 2 meV.

The electric field of the incident circularly polarized light is then included in the Hamiltonian by working in an accelerated reference frame. That is, we substitute $p_x \to p_{x0} + \frac{eE_R}{\omega} \sin(\omega \tau)$ and $p_y \to p_{y0} + \frac{eE_R}{\omega} \cos(\omega \tau)$ where $\tau$ denotes time, e is the electron charge, $\omega$ the angular frequency of the light field and $E_R \approx 0.64 E_0$ is its peak amplitude, taking into account an effective reduction owing to reflections from the substrate, see section S1.2.

---

[*] The next-nearest neighbor tunneling term is left out as it appears as a prefactor to an identity matrix and therefore does not affect the energy difference between the two eigenstates and does not contribute any new terms to the Floquet Hamiltonian.

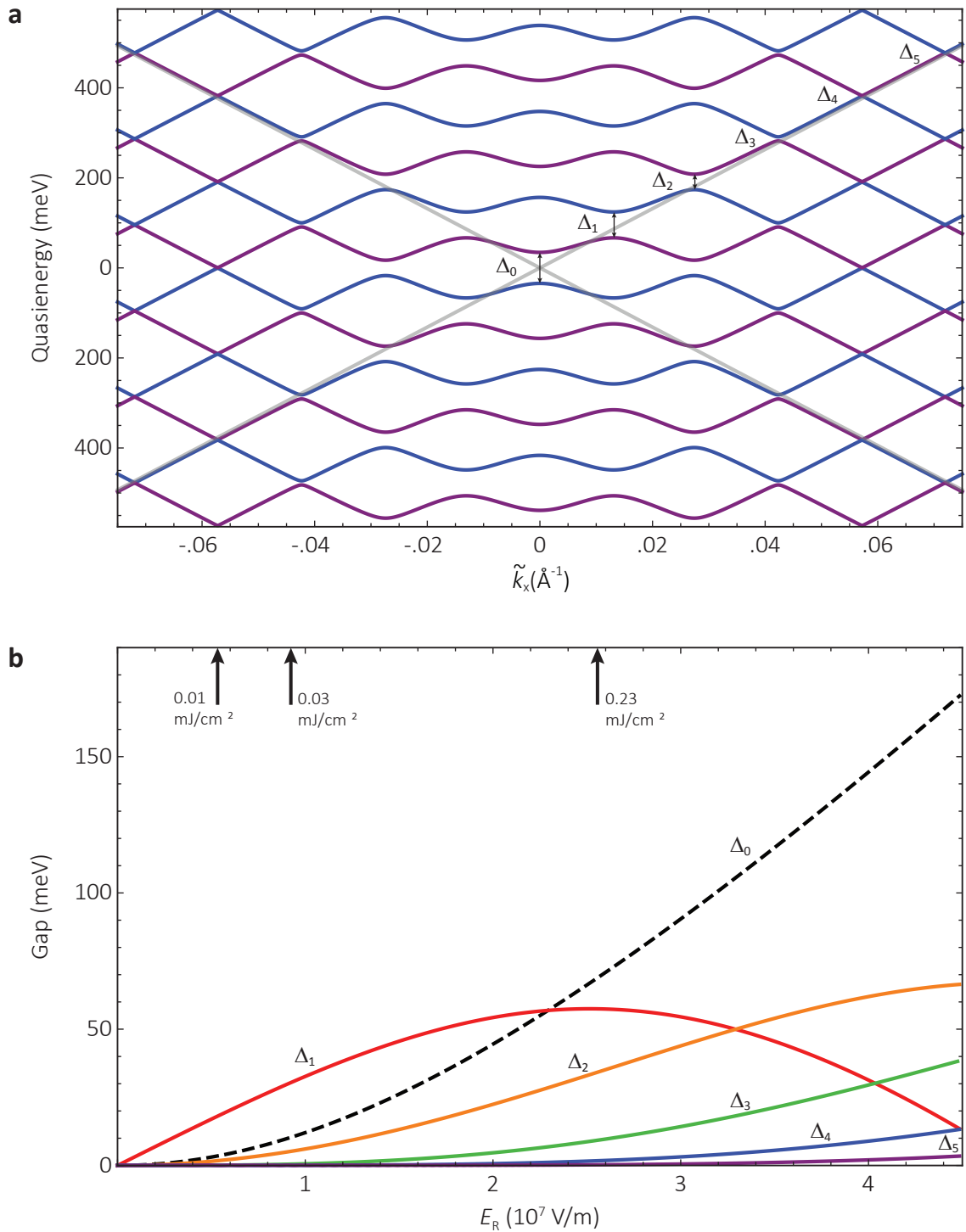

**Fig. S20 | Floquet band structure calculations. a**, Cut through the Floquet band structure near the Dirac point for the maximum fluence used in the main text. The effective field strength, taking into account reflections from the substrate, is $E_R \approx 2.6 \cdot 10^7$ V/m. The equilibrium bands are shown in grey, whilst the two Floquet bands are shown in blue and purple, along with their replicas at $\pm n\hbar\omega$. The light-induced gaps are marked with arrows. **b,** Gap size as a function of effective field amplitude. The amplitudes corresponding to the fluences used in the main text are marked with arrows. The gap at the Dirac point is shown as a black, dashed line, whilst the 1st, 2nd, 3rd, 4th and 5th resonant gaps are red, orange, green, blue and purple, respectively.

In order to compute the Floquet band structure, which describes the behavior of periodically driven Hamiltonians on timescales longer than one driving period[8], we determine the unitary evolution operator for one driving period $T = 2\pi/\omega$. The operator is computed numerically by discretizing the evolution into N steps in real time via

$$\widehat{U}[T,0] = \prod_{j=0}^{N-1} \widehat{U}[(j+1)T/N, jT/N] \approx \prod_{j=0}^{N-1} e^{-i\widehat{H}jT/N\hbar}$$

We find that for $N \approx 100$ our results deviate from the asymptotic limit by less than $10^{-4}$ eV. The Floquet spectrum can then directly be extracted by multiplying $\frac{i\hbar}{T}$ with the logarithm of the eigenvalues of $\widehat{U}[T,0]$ at each $\{p_{x0}, p_{y0}\}$. Owing to the invariance of $\widehat{U}$ upon adding or removing energy offsets of $\hbar\omega$ to the Hamiltonian, the Floquet spectrum is periodic in (quasi)energy. Note that whilst the choice of starting phase for the periodic driving (encoded in the time dependence of $\boldsymbol{p}$) does change $\widehat{U}$ and its eigenstates, the spectrum remains invariant.

Figure S20a shows the Floquet spectrum for the same parameters as the third panel in Fig. 4c of the main text, but with an extended range of momenta and energies and with the quasienergy-periodicity explicitly shown. In the main text, only the Floquet bands closest to the static energy were shown, which is where the non-equilibrium electron population is mostly expected[4,28,31].

In addition to the gap at the Dirac point ($\Delta_0 \approx 69$ meV) and the resonant gaps at $\pm\hbar\omega/2$ ($\Delta_1 \approx 56$ meV), gaps also appear at higher-order resonances. The second-order gap still has a significant size of $\Delta_2 \approx 34$ meV, whilst the third-order gap is only 9 meV large and further gaps are less than 2 meV. This Floquet spectrum was computed using the full tight-binding Hamiltonian $\widehat{H}_{TB}$, but for these parameters the results using $\widehat{H}_D$ differ by less than 3 meV in the range plotted here.

Note that for such strong driving, the Floquet bands are shifted somewhat with respect to the equilibrium band, and hence the positions of the gaps are shifted to smaller momenta than in the weak-driving limit. This effect, related to the Bloch-Siegert shift, originates from the low-frequency limit of the Floquet Hamiltonian, where the Floquet bands are given by the average energy of the equilibrium band that electrons experience during one cycle[45] (unlike in the high-frequency limit, where the average Hamiltonian is experienced). This average energy is higher (lower) than the static energy on the upper (lower) branch of the Dirac cone.

Figure S20b shows how the size of the gaps scales with the effective electric field amplitude $E_R$. The gap at the Dirac point (black, dashed) initially shows a quadratic dependence on the driving field and then closely follows the exact result obtained for the Dirac Hamiltonian[4] of $\Delta_0 = \sqrt{4(v_F e E_R/\omega)^2 + (\hbar\omega)^2} - \hbar\omega$. For the resonant gaps, we find that the scaling is initially linear for the first-order gap, quadratic for the second-order gap, cubic for the third-order gap etc. However, the gaps then saturate at values around 60 meV and subsequently reduce again. The plotted gaps take into account the aforementioned momentum-shift of the gaps by smoothly following the local minima in the band splitting.

A strong driving field is also expected to lead to a phenomenon known as dynamical localization[46], i.e. a reduction of tunnelling in the effective Hamiltonian leading to a flattening of the bands. In the high-frequency limit, this reduction is given by a factor $\mathcal{J}_0(eaE_R/\hbar\omega)$, where $\mathcal{J}_0$ denotes the zeroth-order Bessel function of the first kind. Whilst our calculations do capture this effect when using $\hat{H}_{TB}$ (it is absent for $\hat{H}_D$), we find that it is negligibly small, as expected from the fact that for our parameters this reduction amounts to less than $10^{-4}$.

## Supplementary references